\documentclass[12pt]{article}   	                 
\usepackage{amsmath,amsthm, amssymb,amsfonts,amscd,commath}
\usepackage[colorlinks=true,linkcolor=blue,citecolor=red]{hyperref}
\usepackage{mathrsfs}                             
\usepackage[pdftex]{graphicx}                 
\usepackage{wrapfig}                               
\usepackage[font=footnotesize]{caption}                     
\usepackage{float}
\usepackage{makecell}
\usepackage{multirow}
\usepackage[doublespacing]{setspace}
\usepackage[style=trad-abbrv, backend=bibtex,sorting=none]{biblatex}
\DeclareNameAlias{sortname}{family-given}
\allowdisplaybreaks[4]

\input{./mint.tex}

\usepackage{verbatim}

\DeclareFontFamily{U}{mathx}{\hyphenchar\font45}
\DeclareFontShape{U}{mathx}{m}{n}{
      <5> <6> <7> <8> <9> <10>
      <10.95> <12> <14.4> <17.28> <20.74> <24.88>
      mathx10
      }{}
\DeclareSymbolFont{mathx}{U}{mathx}{m}{n}
\DeclareFontSubstitution{U}{mathx}{m}{n}
\DeclareMathAccent{\widecheck}{0}{mathx}{"71}
\DeclareMathAccent{\wideparen}{0}{mathx}{"75}

\usepackage{scalerel}
\usepackage{stackengine}
\stackMath
\def\hatgap{2pt}
\def\subdown{-2pt}
\newcommand\reallywidehat[2][]{%
\renewcommand\stackalignment{l}%
\stackon[\hatgap]{#2}{%
\stretchto{%
    \scalerel*[\widthof{$#2$}]{\kern-.6pt\bigwedge\kern-.6pt}%
    {\rule[-\textheight/2]{1ex}{\textheight}}
}{0.5ex}
_{\smash{\belowbaseline[\subdown]{\scriptstyle#1}}}%
}}

\numberwithin{equation}{section}
\newtheorem{theorem}{Theorem}[section]
\newtheorem{lemma}[theorem]{Lemma}
\newtheorem{proposition}[theorem]{Proposition}
\newtheorem{definition}[theorem]{Definition}
\newtheorem{corollary}[theorem]{Corollary}
\newtheorem{hypothesis}[theorem]{Hypothesis}
\newtheorem{remark}[theorem]{Remark}

\newenvironment{Proof}%
 {\begin{trivlist} \item[]{\bf Proof. }}%
{\hspace*{\fill}$\rule{.4\baselineskip}{.4\baselineskip}$\end{trivlist}}

\newenvironment{Acknowledgment}%
 {\begin{trivlist}\item[]\textbf{Acknowledgments }}{\end{trivlist}}

\newcommand{\bt}{\begin{theorem}}
\newcommand{\et}{\end{theorem}}
\newcommand{\bl}{\begin{lemma}}
\newcommand{\el}{\end{lemma}}
\newcommand{\bp}{\begin{proposition}}
\newcommand{\ep}{\end{proposition}}
\newcommand{\bd}{\begin{definition}}
\newcommand{\ed}{\end{definition}}
\newcommand{\bc}{\begin{corollary}}
\newcommand{\ec}{\end{corollary}}
\newcommand{\br}{\begin{remark}}
\newcommand{\er}{\end{remark}}
\newcommand{\bh}{\begin{hypothesis}}
\newcommand{\eh}{\end{hypothesis}}
\newtheorem*{theorem*}{Theorem}
\newtheorem*{lemma*}{Lemma}
\newcommand{\be}{\begin{enumerate}}
\newcommand{\ee}{\end{enumerate}}
\newcommand{\beq}{\begin{equation}}
\newcommand{\eeq}{\end{equation}}
\newcommand{\beqs}{\begin{equation*}}
\newcommand{\eeqs}{\end{equation*}}
\newcommand{\bpf}{\begin{Proof}}
\newcommand{\epf}{\end{Proof}}
\newcommand{\bld}{\begin{aligned}}
\newcommand{\eld}{\end{aligned}}
\newcommand{\bhp}{\begin{hypothesis}}
\newcommand{\ehp}{\end{hypothesis}}
\newcommand{\bcs}{\begin{cases}}
\newcommand{\ecs}{\end{cases}}
\newcommand{\R}{\mathbb{R}}
\newcommand{\C}{\mathbb{C}}
\newcommand{\N}{\mathbb{N}}
\newcommand{\Z}{\mathbb{Z}}

\newcommand{\T}{\mathbb{T}}

\newcommand{\rmnum}[1]{\romannumeral #1}
\newcommand{\Rmnum}[1]{\uppercase\expandafter{\romannumeral #1\relax}}

\newcommand{\caO}{\mathcal{O}}

\newcommand{\rmd}{\mathrm{d}}
\newcommand{\rme}{\mathrm{e}}
\newcommand{\rmi}{\mathrm{i}}

\newcommand{\id}{\mathrm{\,Id}}

\renewcommand{\leq}{\leqslant}
\renewcommand{\geq}{\geqslant}

\def\re{\mathop{\mathrm{\,Re}\,}}

\def\rg{\mathop{\mathrm{\,Rg}\,}}

\def\veps{\vepsilon}
\def\veps{\varepsilon}

\def\a{\widetilde{a}}
\def\bnut{\widetilde{\bnu}}
\def\O{\mathcal{O}}
\def\H{\mathcal{H}}
\def\eb{\underline{\widehat{e}_c}}
\def\vsb{\underline{V_s}}
\def\vs{V_s}

\def\x{\mathbf{x}}
\def\k{\mathbf{k}}
\def\bnu{\boldsymbol\nu}

\newcommand{\caB}{\mathcal{B}}

\newcommand{\caL}{\mathcal{L}}
\newcommand{\caT}{\mathcal{T}}
\newcommand{\caN}{\mathcal{N}}

\newcommand{\caF}{\mathcal{F}}

\makeatletter
\newsavebox{\@brx}
\newcommand{\llangle}[1][]{\savebox{\@brx}{\(\m@th{#1\langle}\)}%
  \mathopen{\copy\@brx\kern-0.5\wd\@brx\usebox{\@brx}}}
\newcommand{\rrangle}[1][]{\savebox{\@brx}{\(\m@th{#1\rangle}\)}%
  \mathclose{\copy\@brx\kern-0.5\wd\@brx\usebox{\@brx}}}
\makeatother

\makeatletter
\newcommand{\opnorm}{\@ifstar\@opnorms\@opnorm}
\newcommand{\@opnorms}[1]{%
  \left|\mkern-1.5mu\left|\mkern-1.5mu\left|
   #1
  \right|\mkern-1.5mu\right|\mkern-1.5mu\right|
}
\newcommand{\@opnorm}[2][]{%
  \mathopen{#1|\mkern-1.5mu#1|\mkern-1.5mu#1|}
  #2
  \mathclose{#1|\mkern-1.5mu#1|\mkern-1.5mu#1|}
}
\makeatother

\makeatletter\@addtoreset{figure}{section}\makeatother

\newfam\bifam
\font\tenbi=cmmib10 scaled \magstep1 \font\sevenbi=cmmib10 at 11pt
\font\fivebi=cmmib10 at 6pt \textfont\bifam = \tenbi
\scriptfont\bifam = \sevenbi \scriptscriptfont\bifam= \fivebi

\begin{document}

\title{\centering Weak Diffusive Stability of Roll Solutions at the Zigzag Boundary}
\author{{\centering \scshape{\small Abhijit Chowdhary$\,^\dag$, Mason Haberle$\,^\ddag$, William Ofori-Atta$\, ^\P$ and Qiliang Wu$\,^{\S, *}$}}\\
\textit{\footnotesize\centering  $\dag$. Department of Mathematics, North Carolina State University}\\
\textit{\footnotesize\centering  2311 Stinson Drive, Raleigh, NC 27695, USA}\\
\textit{\footnotesize \centering $\ddag$.  Courant Institute of Mathematical Sciences, New York University, }\\
\textit{\footnotesize\centering  251 Mercer Street, New York, NY 10012, USA}\\
\textit{\footnotesize\centering  $\P,\quad \S$. Department of Mathematics, Ohio University, }\\
\textit{\footnotesize\centering  Morton Hall 321, Athens, OH 45701, USA}\\
\textit{\centerline{\footnotesize $*$. the corresponding author}}
}

\date{\today}
\maketitle
\begin{abstract}
\noindent  
Roll solutions at the zigzag boundary, typically selected by patterns and defects in numerical simulations,
are shown to be nonlinearly stable. This result also serves as an example that linear decay weaker than the classical diffusive decay, together with quadratic nonlinearity, still gives nonlinear stability of spatially periodic patterns. Linear analysis reveals that, instead of the classical $t^{-1}$ diffusive decay rate, small perturbations of roll solutions at the zigzag boundary decay with a $t^{-3/4}$ rate along with time, due to the degeneracy of the quadratic term of the continuation of the translational mode of the linearized operator in the Bloch-Fourier spaces. The nonlinear stability proof is based on a linear decomposition of the neutral translational mode and the faster decaying modes in the Bloch-Fourier space, and a fixed-point argument, demonstrating the irrelevancy of the nonlinear terms.
\end{abstract}


 \hrule
 {\small
 \begin{Acknowledgment}
Qiliang Wu gratefully acknowledges support by the National Science Foundation through grant DMS-1815079. The authors would like to thank Professor Arnd Scheel at University of Minnesota for helpful discussions. 
 \end{Acknowledgment}
  }

\section{Introduction} 
The universal features of patterns arising from both living and nonliving systems, not only in terms of their morphological forms but also of their formation mechanisms, was at first scientifically treated by D'Arcy Thompson in his seminal work on growth and forms \cite{DThompson} in 1917. However, it is not until 1952 when rigorous mathematical analysis was imbued into the field of pattern formation by Alan Turing in his masterpiece entitled ``The Chemical Basis of Morphogenesis" \cite{turing_1952}. Since then, spatially periodic patterns, as one of the most commonly seen patterns in nature, have been intensively studied in various well-established pattern forming systems such as the Ginzburg-Landau equation, the Swift-Hohenberg equation, the Boussinesq equation and many reaction-diffusion models, to just name a few. From the perspective of dynamical systems, the universality of patterns and their dynamics comes from the fact that pattern forming systems in distinctively different settings can be modeled by the same mathematical model, or the same modulation equation for patterns of interest. For the validity of mathematical models of pattern forming systems, it is natural to require that solutions corresponding to patterns observed in the physical system exist and display similar dynamical features as their physical counterparts, among which their stability is a fundamental one, due to the fact that patterns we observe in nature are generically resilient and persistent. Pattern forming systems giving rising to spatially periodic patterns, called roll solutions,
typically accommodate a family of roll solutions parameterized by a continuum of wave numbers. While the wave numbers on the zigzag boundary have been shown to be selected by patterns and their defects in numerical simulations \cite{LloydScheel}, the nonlinear stability of these roll solutions on the zigzag boundary is yet to be proved and thus the topic of this paper.

 
From a broad perspective, we are interested in how the coupling of weak linear stability and nonlinearity affects the nonlinear asymptotic stability of patterns in the setting of smooth dissipative systems. To fix ideas, we investigate the smooth autonomous dynamical system of 
\beq
\label{e:ads}
u_t=F(u),
\eeq
where $u(t;\x)\in\R^N$ is defined on $\R^+\times\R^n$ and $F$ is smooth. Assuming the existence of an equilibrium $u_*$ to \eqref{e:ads}, we study the Lyapunov stability of $u_*$ via the initial value problem 
\beq
\label{e:ads-ivp-u}
\begin{cases}
  u_t=F(u),\\
  u(0)=u_*+v_0,
\end{cases}
\eeq
or equivalently,
\beq
\label{e:adv-ivp-v}
\begin{cases}
  v_t=L v+N(v),\\
  v(0)=v_0,
\end{cases}
\eeq
where 
\[
L:=\frac{\partial F}{\partial u}(u_*), \quad 
N(v)=F(u_*+v)-F(u_*)-Lv.
\]
If the spectrum $\sigma(L)$ lies in the closed left half of the complex plane $\C$; that is, $\sigma(L)\subseteq\{\lambda=a+b\rmi\mid a\leq0, b\in\R\}$, then we say $u_*$ is spectrally stable. 
If $v=0$ is stable in the linearized flow
\[
v_t=Lv,
\]
then we say $u_*$ is linearly stable. Similarly, if $v=0$ is stable in the whole nonlinear flow, then we say $u_*$ is nonlinear stable. In the case when $L$ is hyperbolic and satisfies certain extra condition(s), linear (in)-stability leads to nonlinear (in)-stability. More specifically, if $L$ is of finite dimensional, or, satisfies some regularity condition such as being sectorial \cite{henry}, and every member of the spectrum $\sigma(L)$ admits negative real part (at least one member of $\sigma(L)$ admits positive real part),
then $u_*$ is asymptotically stable (unstable). We note that nonlinear stabilization can take place; see \cite{Gal-Tex-Zum_2017,Rod-Sol_2020} for details.

The nontrivial and most interesting case happens when the spectrum $\sigma(L)$ lies in the left half of the complex plane and touches the imaginary axis.
While the spectrum $\sigma(L)$ can touch the imaginary axis in many ways,  we restrict ourselves to the simple but interesting case when the neutral-stable spectrum $\sigma({L})$ only touches the imaginary axis at the origin; that is,
\[
\sigma({L})\subseteq \{a+b\rmi\mid a\leq0,b\in\R \}, \quad \sigma(L)\cap \rmi\R=\{0\}.
\]
Under suitable assumptions, it is shown that initial perturbations $v_0\in L^1$ are linearly diffusively stable; that is, 
\[
\|v(t,\cdot)\|_{L^\infty}\leq C t^{-N/2}.
\]
Such type of linear diffusive stability is in general not strong enough to ensure higher-order nonlinear decays when $N<3$. However, Schneider showed in his seminal work \cite{schneider_1996}  that the diffusive stability is preserved in the presence of quadratic nonlinear terms, thanks to a cancellation induced by the translational symmetry of the system and a careful estimate of the nonlinear terms via mode-filters. Following his ideas, a cornucopia of work on nonlinear stability of periodic patterns has been developed in the past decades; see \cite{schneider_1998, schneider_1998ARMA, uecker_1999, johnson_2009, johnsonzumbrun_2010,johnsonzumbrun_2011sj,johnsonzumbrun_2011,johnsonzumbrunpascal_2011, SSSU_12, GSWZ_18, HRS_2020}.
While most of previous studies focus on the case of the preservation of diffusive stability in the presence of various nonlinear terms, we instead focus on the effect of weakening in the linear diffusive decay on the nonlinear dynamics of patterns; \cite{GSWZ_18,HRS_2020} for the recent work in this direction.
  
\paragraph{Heat equation: Relevancy and irrelevancy of nonlinear terms} To further illustrate the ideas, we look at the nonlinear heat equation,
\beq
\label{e:nlheat}
u_t=\Delta_\x u+f(u),
\eeq
where $u(t,\x)\in\R$ with $(t,\x)\in\R^+\times\R^n$ and $\x=(x_1,\cdots, x_n)$, and $f(0)=f^\prime(0)=0$. The equilibrium $u\equiv0$ is linearly stable. More specifically, we have that
\[
\sigma(\Delta_\x)=(-\infty,0],
\]
and the linear heat equation $u_t=\Delta_\x u$ admits the Gaussian decay estimates
\[
\|\partial_\x^\alpha u(t,\cdot)\|_{L^p(\R^n)}\leq Ct^{-[\frac{n}{2}(\frac{1}{q}-\frac{1}{p})+\frac{|\alpha|}{2}]}\|u(0,\cdot)\|_{L^q(\R^n)},
\]
where $1\leq q\leq p\leq \infty$ and $\alpha=(\alpha_1,\cdots, \alpha_n)\in\N^n$ is the multi-index of partial derivatives with $|\alpha|=\displaystyle \sum_{i=1}^n\alpha_i$. In particular, for $p=\infty$, $q=1$ and $\alpha=0$, the decay estimate reduces to
\beq
\label{e:heat-lin-est}
\|u(t,\cdot)\|_{L^\infty(\R^n)}\leq Ct^{-\frac{n}{2}}\|u(0,\cdot)\|_{L^1(\R^n)},
\eeq
which we refer to as the \textit{diffusive decay estimate} and the algebraic decay rate $t^{-n/2}$ is called the \textit{diffusive decay rate}.

The stability of the equilibrium $u=0$ in the nonlinear case \eqref{e:nlheat}, however, depends on the type of nonlinearity $f(u)$.
Intuitively, assuming $L^1$ initial data, we can exploit the Gaussian estimates \eqref{e:heat-lin-est} to determine whether the diffusion term $\Delta_\x u$ or the nonlinear term $f(u)$ is dominant in terms of their temporal decay rates, leading to the classification of nonlinear terms into relevant, irrelevant and critical \cite{Fujita_1966,Hayakawa_1973,Levine_1990,Bri-Kup-1994-2,Den-Lev_2000}. More explicitly, we have 
\[
\|\Delta_\x u\|_{L^\infty(\R^n)}\sim t^{-(\frac{n}{2}+1)}, \quad \|f(u)\|_{L^\infty(\R^n)}\sim t^{-k},
\]
where the later estimate is derived based on the Gaussian estimates \eqref{e:heat-lin-est} on $u$ and its derivatives. The nonlinear term $f$ is called \textit{irrelevant} if $k>n/2+1$, \textit{relevant} if $k<n/2+1$, and \textit{critical} if $k=n/2+1$. We expect that, given any irrelevant nonlinear term, the equilibrium $u=0$ is nonlinearly stable with the same diffusive decay rate as the linear case; any relevant nonlinear term makes the local nonlinear dynamics near the equilibrium $u=0$ different from its linear counterpart; the case for the critical one is undetermined and typically needs to be handled on a case-by-case basis.
For example, we let $f(u)=u^m$ and the application of Gaussian estimates leads to
\[
\|f(u)\|_{L^\infty(\R^n)}=
\|u^m\|_{L^\infty(\R^n)}=\|u\|_{L^\infty(\R^n)}^m\sim t^{-\frac{mn}{2}},
\]
which implies that $f(u)=u^m$ is irrelevant if $m>1+2/n$, relevant if $m<1+2/n$ and critical if $m=1+2/n$; see Table \ref{t:non-cls} for more examples. Indeed, Fujita showed in 1966 \cite{Fujita_1966} that, if $m<1+2/n$, then the solution $u$ blows up in finite time; if $m>1+2/n$, then $u=0$ is asymptotically stable with the decay rate $t^{-n/2}$. The critical case $m=1+2/n$ also admits finite-time blow-up, according to the work by Hayakawa \cite{Hayakawa_1973}. 
\begin{table}[ht]
    \centering
\begin{tabular}{|c|c|c|}
\hline
 Classification & Example & Dynamics\\
\hline
Irrelevant ($k>n/2+1$) & $f=u \Delta u, |\nabla u|^2, u^m (m>1+2/n) $ & $\|u\|_{L^\infty}\sim t^{-n/2}$
\\
\hline
Critical ($k=n/2+1$) & $f=\pm u^{1+2/n}$ & Undetermined \\
\hline
 Relevant ($k<n/2+1$) & $f=u^m (m<1+2/n)$ & Finite time blow up\\
\hline
\end{tabular}
\caption{Examples of irrelevant, critical, relevant nonlinear terms in the scalar heat equation}
\label{t:non-cls}
\end{table}

\begin{figure}
    \centering
    \includegraphics[scale=0.3]{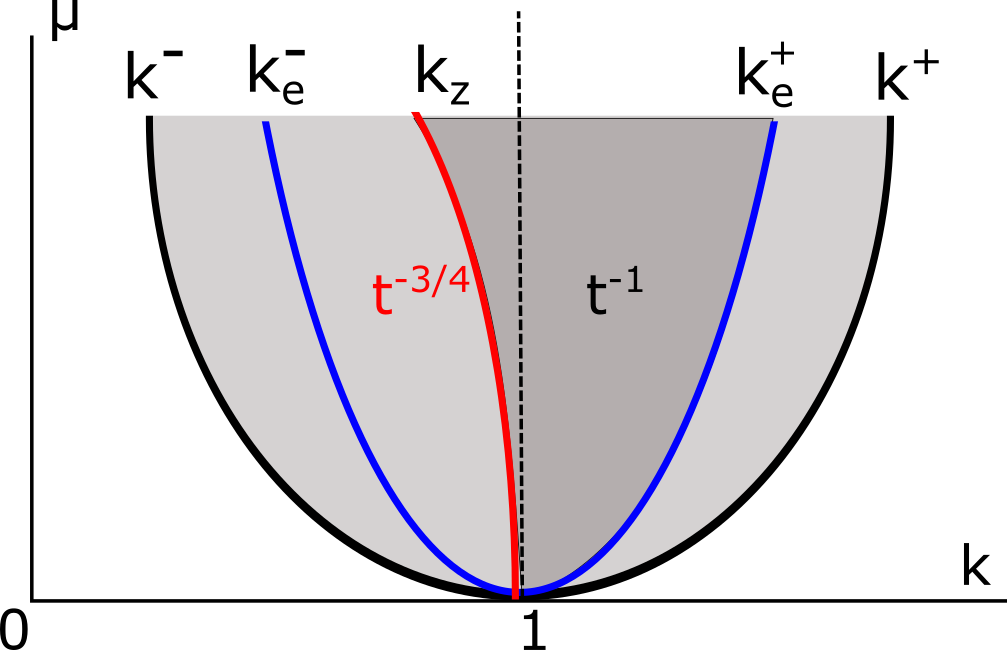}
    \caption{The illustration of the Busse balloon for roll solutions in the 2D SHE.}
    \label{f:balloon}
\end{figure}

\paragraph{Spatially periodic patterns: Previous results and open questions}
In spite of not being rigorous, this intuitive classification of nonlinear terms still works in the study of spatially periodic patterns, not directly but in a more subtle way, which typically involves a proper change of coordinates based on a decomposition of the neutral modes and stable modes. To fix ideas, we study the prototypical model of spatially periodic patterns--the isotropic Swift-Hohenberg Equation (SHE)
\begin{equation} \label{e:she}
  u_t = -(1 + \Delta_\x)^2u + \mu u - u^3
\end{equation}
where $u(t,\x)$ is a real-valued function defined on  $[0,\infty)\times\R^n$ and $\mu\in\R$ is the bifurcation parameter. The homogeneous equilibrium $u\equiv0$ is stable for $\mu<0$ and unstable for $\mu>0$. For $0<\mu\ll 1$, the instability of the homogeneous equilibrium gives rise to a family of even spatially periodic solutions, called roll solutions. More specifically, setting $\mu>0$ for the rest of the paper and denoting $\veps:=\sqrt{\mu}$, we have the following lemma from \cite{Col-Eck_1990, Mielke95,mielke_1997}.
\bl[Existence of roll solutions]
\label{l:ex-roll}
There exists $0<\veps_0\ll1$ such that for any $\veps\in(0,\veps_0)$ and the wave number $k\in(k^-,k^+)$ with $k^\pm=\sqrt{1\pm\veps}$, the stationary rescaled one-dimensional SHE,
\[
-(1+k^2\partial_\xi^2)^2u+\veps^2 u-u^3=0,
\]
admits a unique roll solution $u_p(\xi;k)$ which is $2\pi$-periodic and even in $\xi$ with $u_p(0;k)>0$; see Figure \ref{f:balloon}. The roll solution admits the property $u_p(\xi+\pi;k)=-u_p(\xi;k)$ and the leading order expansion
\beq
\label{e:roll-exp}
u_p(\xi) = a_1\cos({\xi}) + a_3\cos{(3\xi)} + \O(\a^5),
\eeq
where
\[
\begin{aligned}
   a_1 &= \textstyle{\frac{1}{\pi}\int_0^{2\pi} u_p(\xi) \cos({\xi}) d\xi} 
     = \a + \a^3/512 + \O(\a^4), \\
a_3 &= \textstyle{\frac{1}{\pi}\int_0^{2\pi} u_p(\xi) \cos{(3\xi)} d\xi} 
     = -\a^3/256 + \O(\a^4),\\
     \widetilde{a}&=\sqrt{\frac{4[\veps^2-(k^2-1)^2]}{3}}.
\end{aligned}
\]
\el
\begin{remark}
We note that the symmetric property $u_p(\xi+\pi;k)=-u_p(\xi;k)$ results from the persistence of translation symmetry $u(\xi)\to u(\xi+\xi_0)$ and the reflection symmetry $u\to-u$ in the construction of roll solutions via the Lyapunov-Schmidt reduction.
\end{remark}
The rotational and translational symmetries of the system \eqref{e:she} guarantee that 
\[
u_p(\k\cdot\x+\phi;|\k|)
\]
for any $\phi\in\R$, $\k=(k_1,\cdots,k_n)\in\R^n$ with $|\k|=\sqrt{\sum_{i=1}^nk_i^2}\in(k^-,k^+)$ is also a roll solution. 

Eckhaus first discovered in 1965 that not all roll solutions are \textit{spectrally} stable, due to sideband instability induced by perturbations with period close to, but not equal to the period of roll solutions \cite{Eckhaus_1965}. 
In particular, for one-dimensional SHE,  $n=1$, the roll solution $u_p$ is \textit{spectrally} stable for $k\in (k_e^-,k_e^+)$ with $k_e^{\pm}=1 \pm \frac{\veps}{\sqrt{12}}+\caO(\mu)$; see Figure \ref{f:balloon}. More specifically, the continuation of the eigenvalue $0$ in the Fourier space admits the expansion
\[
\lambda(\nu_1;k)=-a_{20}(k)\nu_1^2+\caO(\nu_1^3),
\]
where $\nu_1$ is the Fourier wave number and $a_{20}(k)>0$ if and only if $k\in(k_e^-,k_e^+)$. 
The rigorous nonlinear stability proof was first given much latter by Schneider in 1995 via renormalization techniques \cite{schneider_1996}, where the perturbation $v=u-u_p$ was shown to decay diffusively as time goes to infinity; that is,
\[
\|v\|_{L^\infty(\R)}\sim t^{-1/2}.
\]
For two-dimensional SHE, $n=2$, the roll solutions are also subject to another secondary instability, called zigzag instability, caused by perturbations along the transversal direction of roll solutions \cite{mielke_1997}. More specifically, the continuation of eigenvalue $0$ now admits the expansion
\[
\lambda(\bnu;k)=-a_{20}(k)\nu_1^2-a_{02}(k)\nu_2^2+\caO(|\bnu|^3),
\]
where $\bnu:=(\nu_1,\nu_2)$ is the Fourier wave vector and $a_{02}(k)>0$ if and only if $k>k_z$ with $k_z=1-\frac{\veps^4}{512}+\caO(\veps^5)$. A proof of the nonlinear stability of roll solutions in the two-dimensional SHE was given by
Uecker in 1999 via renormalization techniques \cite{uecker_1999} and the decay rate of the perturbation $v=u-u_p$ is again diffusive; that is,
\[
\|v\|_{L^\infty(\R^2)}\sim t^{-1}.
\]
Parallel and generalized results in Ginzburg-Landau equation were obtained also in the 1990s \cite{CEE92,Bri-Kup_1992} and later in reaction-diffusion systems \cite{johnsonzumbrun_2011,SW_15,SSSU_12} and viscous conservation laws \cite{johnson_2009,johnsonzumbrun_2010,johnsonzumbrun_2011sj,johnsonzumbrunpascal_2011}. 

More recently, Guillod \textit{et al.} \cite{GSWZ_18} showed that spatially periodic solutions on the Eckhaus boundary in the Ginzburg-Landau equation is nonlinearly stable. Putting this result in the same context as the previous ones, the continuation of the eigenvalue $0$ in \cite{GSWZ_18} take the expansion
\[
\lambda(\nu_1)=-a_{04}\nu_1^4+\caO(\nu_1^5),
\]
and the perturbation $v$ decays algebraically with decay rate $t^{-1/4}$ as time goes to infinity; that is,
\[
\|v\|_{L^\infty(\R)}\sim t^{-1/4}.
\]
This particular work gives an example that the coupling of a weakened linear stability and a quadratic nonlinearity still lead to nonlinear stability of spatial periodic patterns, which leads naturally to the following interesting open questions.
\begin{itemize}
    \item  Can the nonlinear stability of spatially periodic patterns always hold while we keep on weakening the linear stability? If not, what is the critical decay rate of the linear stability beyond which the nonlinear stability does not hold anymore? Can we describe the local dynamics of spatially periodic patterns beyond the critical linear decay?
    \item Within the regime where the weakening of linear stability still leads to nonlinear stability of patterns, is there a general theorem establishing the relation between the leading order expansion of the eigenvalue $0$ and the nonlinear decay rate? More specifically, we recall the abstract dynamical system \eqref{e:ads}, $u_t = F(u)$ which admits a spatially periodic equilibrium $u_*$ solving $F(u_*)=0$. We also recall that the perturbation  $v=u-u_*$ yields the perturbed system \eqref{e:adv-ivp-v}, $v_t=L v+N(v)$, 
where we assume that the initial perturbation $v_0$ is small in the $L^1$ norm sense, and the spectrum $\sigma (L)$  lies in the left half of the complex plane and touches the imaginary axis only at the origin. In addition, the continuation of the eigenvalue $0$ in the Bloch-Fourier space admits the expansion
\[
\lambda(\bnu)= - \sum_{i=1}^{n} a_{i(2m_i)}\nu_{i}^{2m_{i}} + h.o.t ,
\]
where  $\bnu=(\nu_1,\cdots,\nu_n) $ is the wave vector, $a_{i(2m_i)}>0$ and $m_i\in\Z^+$ for all $i=1,\cdots, n$. Such an expansion yields a weak version of diffusive decay in the linear flow via the neutral part of the kernel $\mathrm{e}^{\lambda(\bnu)t}$ in the Bloch-Fourier space, which in turn leads to $\|\widehat{v}(t,\cdot)\|_{L^1}\sim t^ {-\sum_{i=1}^{n}1/(2m_i)}$ in the linear flow. We are interested in whether such a weak version of diffusive decay of the perturbation $v$ is preserved in the presence of nonlinear terms; that is,
\[
\|v(t,\cdot)\|_{L^\infty(\R^n)}\sim t^ {-\sum_{i=1}^{n}\frac{1}{2m_i}}.
\]
\end{itemize}

\paragraph{Main result}In this work, we do not pretend to give ultimate answers to the above open questions, but rather provide another example where nonlinear stability holds under weakened linear stability,
enriching the example pool. More specifically, we apply techniques similar to \cite{GSWZ_18} and \cite{MSU01} to prove the nonlinear stability of roll solutions of the real two-dimensional Swift-Hohenberg equation \eqref{e:she} on the zigzag boundary.
\begin{theorem} \label{thm:main}
For any $0<\mu=\veps^2\ll1$, $u_p(k_zx_1;k_z)$ is nonlinearly stable in the two-dimensional SHE \eqref{e:she}; that is, there exists $\delta > 0$ such that for any initial perturbation $v_0(\x):=u(0,\x)-u_p(k_zx_1;k_z)
  \in L^2(\R^2)$, satisfying 
  \[
  \norm{\widehat{v_0}}_{L^1(\R^2)}+\norm{\widehat{v_0}}_{L^\infty(\R^2)} \leq\delta,
  \]
  where $\widehat{v_0}$ represents the Fourier transform of $v_0$,  
  the $L^\infty$-norm of the perturbation $v(t,\x)=u(t,\x)-u_p(k_zx_1;k_z)$ goes to zero as time goes to infinity. More specifically, there exists $C>0$ such that
  \[
   \norm{v(t,\cdot)}_{L^\infty(\R^2)} \leq
    C\frac{\norm{\widehat{v_0}}_{L^1(\R^2)}+\norm{\widehat{v_0}}_{L^\infty(\R^2)}}{(1+t)^{3/4}}, \quad \forall t >0 .
  \]
\end{theorem}
\begin{remark}
We note that with
minor tweaks we can then extend our argument from the boundary case $k=k_z$ to the generic case $k\in(k_z, k_e^+)$ and provide an alternate
proof for the already widely known result \cite{uecker_1999}, while the given scheme seems to be not sufficient to recover the nonlinear stability result in the 1D case \cite{schneider_1996}; see also Remark \ref{r:gen} for a more detailed comment.
\end{remark}

\begin{remark}
It is of interest and requires further study whether the modulation dynamics just above the zig-zag boundary, given by a Cahn-Hilliard type equation \cite{CCCG_2002, hoyle_2006, dull_2013}, behaves similar to the full equation; see \cite{schneider_1998ARMA} for such a discussion in a different setting.
\end{remark}
The rest of the paper is organized as follows. In Section \ref{s:2}, we introduce the Bloch-Fourier transform and the mode filter decomposition, where the irrelevancy of nonlinear terms can be observed intuitively as in the case of the nonlinear scalar heat equation. We then give the rigorous proof of Theorem \ref{thm:main} in Section \ref{s:3} via a contraction mapping argument on the variation of constant formula posed on a fine-tuned space. For clarity and conciseness, we relegate to the appendix the calculation of leading order expansion of the eigenvalue $0$ continuation, sectorial properties of $\caL_p$ in the Bloch-Fourier space, estimates of various secondary nonlinear terms and a brief proof for the generic case $k\in(k_z, k_e^+)$.

\paragraph{Notation} Throughout we will use the following notation.
\begin{itemize}

 \item The inner product on $\R^2$ is denoted as  $\x\cdot\mathbf{y}:=\sum_{j=1}^2x_jy_j$, for all $\x=(x_1,x_2),\mathbf{y}=(y_1,y_2)\in\R^2$.
 \item The inner product on the Hilbert space $L^2(\T_{2\pi})$ is denoted as $\langle u,v\rangle:=\int_{\T_{2\pi}}u(\xi)\bar{v}(\xi)\rmd \xi$, for any  $u,v\in L^2(\T_{2\pi})$,
          where $\bar{v}$ denotes the complex conjugate of $v$.
 \item The inner products on $\ell^2$, or the $\ell^p$--$\ell^q$ pairing, is denoted as $\llangle\underline{u},\underline{v}\rrangle:= \sum_{j\in\Z}{u}_j\bar{v}_j$, for any $\underline{u}=\{u_j\}_{j\in\Z}\in\ell^p,\underline{v}=\{v_j\}_{j\in\Z}\in\ell^q$  with $\frac{1}{p}+\frac{1}{q}=1, 1\leq p, q\leq \infty$.
\item  For $p\in[1,\infty)$, $n\in\N$, we define the discrete Sobolev space $w^{n,p} := \left\lbrace\underline{u}  \;\middle|\;\norm{u}_{w^{n,p}}<\infty \right\rbrace$  where the Sobolev norm takes the form $\norm{u}_{w^{n,p}}:=\bigg( \sum_{i=0}^{n}\Big(\sum_{j\in \Z}|j^i u_j|^p\Big)  \bigg)^{\frac{1}{p}}$;
while for $p=\infty$, $n\in\N$, we have $w^{n,\infty} := \left\lbrace\underline{u}  \;\middle|\;\norm{u}_{w^{n,\infty}}<\infty \right\rbrace$  where
$\norm{u}_{w^{n,\infty}}:= \max_{i=0,\cdots, n}\left(\sup_{j\in \Z}|j^i u_j|\right)$.
We note that $w^{0,p}=\ell^p$.
\item For any $u\in L^2(\R^2)$, we use the notations $\caF u$ and $\widehat{u}$ interchangeably for its Fourier transform, and $\caF^{-1}u$ and $\widecheck{u}$ for its inverse Fourier transform; that is,
\[
(\caF u)(\bnu)=\widehat{u}(\bnu):=\frac{1}{(2\pi)^2}\int_{\R^2} u(\x)\rme^{-\rmi \x\cdot\bnu}\rmd \x; \qquad
(\caF^{-1 }u)(\bnu)=\widecheck{u}(\bnu):=\int_{\R^2} u(\x)\rme^{\rmi \x\cdot\bnu}\rmd \x.
\]
\item  For $u\in L^2(\T_{2\pi})$, we use the notations $\caF_d u$ and $\underline{\widehat{u}}$ interchangeably for its Fourier series; that is, 
\[
(\caF_d u)_j=\widehat{u}_j:=\frac{1}{2\pi}\int_{\T_{2\pi}}u(\xi)\rme^{-\rmi j \xi}\rmd \xi.
\]
\item The convolution of two functions $u, v: X\to\C$ is defined as $u*v(x):=\int_X u(x-\widetilde{x})v(\widetilde{x})\rmd \widetilde{x}$,
where we use the Lebesgue measure if $X$ is Euclidean and the counting measure if $X$ is discrete.
In addition, we denote 
\[u^{*n}:=\overbrace{u*\cdots*u}^{n \text{ of } u}.
\]
\end{itemize}
We denote the Euclidean norm in Euclidean spaces as $|\cdot|$, the norm in a general Banach space $\mathscr{X}$
as $\|\cdot\|_{\mathscr{X}}$, and the norm of a linear operator from a Banach space $\mathscr{X}$ to $\mathscr{Y}$ as
$\opnorm{\cdot}_{\mathscr{X}\rightarrow\mathscr{Y}}$. For the case $\mathscr{Y}=\mathscr{X}$, the last norm notation simply becomes
$\opnorm{\cdot}_{\mathscr{X}}$. For $\mathscr{X}=L^p(\R^2), L^p(\T_{2\pi}), L^p(\T_1\times\R)$, or $\ell^p$, the second norm notation simply becomes $\|\cdot\|_p$, if there is no ambiguity. At last, we use the universal notation $C$ for positive constants throughout the paper.

\section{Nonlinear irrelevancy under the mode filter decomposition} 
\label{s:2}
In this section, we split the SHE into the neutral modes and stable modes via a mode filter decomposition in the discrete Bloch-Fourier space. Moreover, we also give refined linear estimates and detailed expressions of nonlinear terms under such a decomposition, from which the irrelevancy of nonlinear terms are followed via the intuitive counting of decay rates analogous to the heat equation case. A rigorous proof of the nonlinear irrelevancy will be given in the next section.

\subsection{Mode filter decomposition in the discrete Bloch-Fourier space} 
\label{ss:2.1}
From now on, we fix 
\[
n=2, \quad 0<\veps<\veps_0, \quad k\in(k^-,k^+),\quad \x=(x_1,x_2), \quad \bnu=(\nu_1,\nu_2),
\]
denote 
\beq
\kappa:=k^2-1, 
\eeq
introduce the rescaling $x_1\to kx_1$,
and study the initial value problem of the rescaled SHE,
\beq
\label{e:rs-SHE}
\begin{cases}
  u_t=-\left(1+(1+\kappa)\partial_{x_1}^2+\partial_{x_2}^2\right)^2u+\veps^2 u-u^3,\\
  u(0,\x)=u_p(x_1)+v_0(\x),
\end{cases}
\eeq
or equivalently, the perturbation equation of $v(t,\x):=u(t,\x)-u_p(x_1)$,
\beq
\label{e:SHE-ivp-v}
\begin{cases}
   v_t =\caL_p v + \caN_p(v), \\
   v(0,\x)=v_0(\x),
\end{cases}
\eeq
where 
\beq
\label{e:L&N}
 \caL_p v:= -\left(1+(1+\kappa)\partial_{x_1}^2+\partial_{x_2}^2\right)^2v+\veps^2 v-3u_p^2v, \quad \caN_p(v)=-3u_pv^2-v^3.
\eeq
We first use the Bloch-Fourier transform to study the spectrum of $\caL_p$, especially the leading order expansion of the eigenvalue $0$ continuation.

\paragraph{Bloch-Fourier transform} The linearized operator of the stationary SHE at the roll pattern $u_p$,
\beq
\label{e:lin-roll}
\begin{matrix}
\caL_p: & H^4(\R^2) & \longrightarrow & L^2(\R^2)\\
& v & \longmapsto &  -\left(1+(1+\kappa)\partial_{x_1}^2+\partial_{x_2}^2\right)^2v+\veps^2 v-3u_p^2v
\end{matrix}
\eeq
as a differential operator admits coefficients that are $2\pi$-periodic in $x_1$ and constant in $x_2$. It is well-known that the spectral analysis of constant-coefficient differential operators can be readily done via the Fourier transform, while the spectral analysis of periodic-coefficient differential operators can be achieved via the Bloch wave decomposition \cite{reedsimon78}. To analyze the spectrum of $\caL_p$, we denote $\T_k = \R/k\Z$ and introduce the Bloch-Fourier transform
\beq
\label{e:bft}
\begin{matrix}
\caB :& L^2(\R^2) &\longmapsto& L^2(\T_1 \times \R, L^2(\T_{2\pi}))\\
& v & \longrightarrow & \caB v(\bnu, \xi)  = \sum_{j \in \mathbb{Z}} \widehat{v}(\nu_1 + j, \nu_2)\rme^{\rmi j\xi},
\end{matrix}
\eeq
We note that the Bloch-Fourier transform $\caB$ is an isomorphism \cite{reedsimon78} and the linearized operator $\caL_p$ is block-diagonalized on the Bloch-Fourier space; that is, $\widehat{\caL_p}:=\caB\circ\caL_p\circ\caB^{-1}$ admits the direct integral form
\[
\widehat{\caL_p}=\int_{\T_1\times\R^1}\widehat{\caL_p}(\bnu)\rmd \bnu,
\]
where 
\beq
\label{e:lin-bft-nu}
\begin{matrix}
\widehat{\caL_p}(\bnu) :&  H^4(\T_{2\pi}) &\longrightarrow & L^2(\T_{2\pi})\\
&w(\xi) &\longmapsto & -\left(1 + (1 + \kappa)(\partial_\xi + i\nu_1)^2 - \nu_2^2\right)^2w + \veps^2w - 3u_p^2(\xi)w.
\end{matrix}
\eeq
We now introduce a proposition about the spectrum of the linear operators $\caL_p$, $\widehat{\caL_p}$ and $\widehat{\caL_p}(\bnu)$.
\begin{proposition}[Spectral stability]
\label{p:spectral}
There is a sufficiently small $\veps_1\in(0, \veps_0)$ such that for any fixed $\veps\in(0,\veps_1)$, the operator $\caL_p$ admits the following spectral properties.
\begin{enumerate}
    \item $\displaystyle\sigma(\caL_p)=\sigma(\widehat{\caL_p})=\bigcup_{\bnu\in\T_1\times\R} \sigma(\widehat{\caL_p}(\bnu))\subseteq \R$.
    \item There exist $k_z,k_e^+\in(k^-,k^+)$ so that  $\displaystyle\sigma(\caL_p)\subseteq (-\infty,0]$ if  and only if $k=\sqrt{1+\kappa}\in[k_z,k_e^+]$.
    \item $0$ is a simple eigenvalue of $\widehat{\caL_p}(0)$ with $\displaystyle e_0:=\frac{u_p^\prime}{\|u_p^\prime\|_2}$ as its associated normalized eigenfunction. Moreover, there exists $r_0>0$ such that the eigenpair $(0, e_0)$ admits the unique analytic continuation $(\lambda(\bnu),e(\bnu;\xi))$  for $|\bnu|<r_0$ and $e(\bnu;\xi)$ normalized, satisfying $\widehat{\caL_p}(\bnu)e(\bnu)=\lambda(\bnu)e(\bnu)$ with 
    \beq \label{eq:spec}
    \langle e(\bnu;\cdot)-e_0, e_0\rangle=0, \quad e(\bnu)-e_0=\caO(|\bnu|), \quad e(\bnu)=e_r(\bnu)+\rmi \nu_1 e_i(\bnu),
    \eeq
    where $e_r(\bnu)$ is an odd real-valued function and $e_i(\bnu)$ is an even real-valued function.
    Moreover, we have
    \beq \label{eq:spec2}
     \lambda(\bnu,\veps^2, \kappa)=a_{20}(\veps^2, \kappa){\nu_1}^2
  + a_{02}(\veps^2,\kappa){\nu_2}^2 + a_{04}(\veps^2, \kappa){\nu_2}^4+\caO(\nu_1^4+\nu_2^6), 
    \eeq
    where 
    \begin{enumerate}
        \item $a_{20}, a_{02}<0$ if and only if $k
 \in(k_z,k_e^+)$; or, equivalently, $\kappa\in (k_z^2-1, (k_e^+)^2-1)$;
        \item For $k=k_z$, $a_{20}=-4+\O(\veps^3)<0$, $a_{02}=0$, $a_{04}=-1+\O(\veps^4)<0$;
        \item For $k=k_e^+$,
        $a_{20}=0$, $a_{02}<0$.
    \end{enumerate}
\end{enumerate}
\end{proposition}
\begin{proof} Mielke gave a proof of this proposition in his work \cite[Theorem 3.2]{mielke_1997}, except for that he did the expansion of the eigenvalue $\lambda(\bnu)$ only up to the quadratic order. However, the expansion of $\lambda(\bnu)$ up to the quartic order can be readily derived by plugging $k_z(\veps)=-\frac{\veps^2}{512}+\caO(\veps^5)$ into 
\[
\lambda(\bnu)=\frac{-n_1-\sqrt{n_1^2-4n_0}}{2},
\]
where $n_1$ and $n_0$ are defined in \cite[Line 12, Page 844]{mielke_1997}. A self-contained proof is also available \href{https://drive.google.com/file/d/1rfRzUX7aXUooEilc7UGG8PfOCFWblf05/view}{here} for interested readers.
\end{proof}

\paragraph{Intuition on nonlinear relevancy and irrelevancy} We recall the SHE in terms of the perturbation $v$ in \eqref{e:SHE-ivp-v}; that is,
\[
v_t =\caL_p v + \caN_p(v),
\]
where, based on its spectral properties in Proposition \ref{p:spectral}, the linearized operator $\caL_p: H^4(\R^2)\to L^2(\R^2)$ is a generalized Laplacian. More specifically, we have the following analogies.
\begin{center}
\begin{tabular}{|c|c|c|c|}
\hline
Equation  & Bloch-Fourier & Continuation of $\lambda=0$ & Decay \\
\hline
$v_t=(\partial_x^2+\partial_y^2) v$  &$(\caB v)_t=\left[(\partial_\xi+\rmi\nu_1)^2-\nu_2^2\right]\caB v$ &$\lambda=-\nu_1^2-\nu_2^2$ & \multirow{2}{*}[-0.1in]{$\|u\|_{L^\infty}\sim t^{-1}$}
\\ \cline{1-3}
\makecell{$v_t=\caL_p v$, \\ $\kappa\in (\kappa_z,\kappa_e^+)$ } &$(\caB v)_t=\widehat{\caL_p}\caB v$ &$\lambda=-a_{20}\nu_1^2-a_{02}\nu_2^2+\caO(|\bnu|^4)$ &  
\\
\hline
\hline
$v_t=(\partial_x^2+\partial_y^4) v$ & $(\caB v)_t=\left[(\partial_\xi+\rmi\nu_1)^2-\nu_2^4\right]\caB v$ & $\lambda=-\nu_1^2-\nu_2^4$&  \multirow{2}{*}[-0.1in]{$\|u\|_{L^\infty}\sim t^{-3/4}$}
\\ \cline{1-3}
\makecell{$v_t=\caL_p v$, \\ $\kappa= \kappa_z$ } &$(\caB v)_t=\widehat{\caL_p}\caB v$ &$\lambda=-a_{20}\nu_1^2-a_{04}\nu_2^4+\caO(|\bnu|^4)$ &  
\\
\hline
\end{tabular}
\end{center}
Moreover, we exploit the intuition we derive from the heat equation to evaluate the temporal decay rates of both linear and nonlinear terms via the linearized flow; that is,
\[
\|\caL_p v\|_{L^\infty}\sim t^{-7/4}, \quad \|N_p(v)\|_{L^\infty}=\|-3u_pv^2-v^3\|_{L^\infty}\sim t^{-3/2},
\]
which misleadingly indicates that the nonlinear terms are relevant. This false conclusion results from the fact that we applied our intuitive reasoning on $v$, the sum of both neutral and stable modes, instead of the neutral modes. As a result, the remedy here is to study the system in a refined coordinate system where the neutral and stable modes are properly separated via the mode filter decomposition; see \cite{GSWZ_18} for a similar analysis in the Ginzburg-Landau equation.

\paragraph{SHE in the discrete Bloch-Fourier space} In the Bloch-Fourier space, the SHE \eqref{e:SHE-ivp-v} in terms of the perturbation $v$ takes the form 
\begin{equation}
\label{e:bloch-SHE}
   V_t = \widehat{\caL_p}(\veps^2, \kappa, \bnu)V - 3 u_pV^{* 2} - V^{*3},
\end{equation}
where we introduced $V(t,\bnu):=\caB v(t,\bnu)$ and exploit the fact that, for $u \in L^2(T_{2\pi}), v_1, v_2 \in L^2(\mathbb{R}^2)$, 
\beq
\label{e:BF-con}
\caB (uv_1) = u\caB(v_1), \qquad \caB(v_1v_2) = \caB v_1 * \caB v_2 ,
\eeq
where the function $(uv_1)(\x)=u(x_1)v_1(\x)$; see Lemma 8.1 in \cite{schneider_1998ARMA} for the proof of \eqref{e:BF-con}.

It is typically inevitable to go beyond the $L^2$ space to general $L^p$ space in order to perform proper analysis on nonlinear terms. Noting that 
\[
\|\widehat{v}\|_{L^p(\R^2)}^p=\int_{\R^2}|\widehat{v}(\bnu)|^p\rmd \bnu=\int_{\T_1\times\R}\left(\sum_{j\in\Z}|\widehat{v}(\nu_1+j,\nu_2)|^p\right) \rmd \bnu=\int_{\T_1\times\R}\|\caF_d\caB v\|_{\ell^p}^p\rmd \bnu,
\]
we see that it is more proper to work in the discrete Bloch-Fourier space $L^p(\T_1\times\R,\ell^p)$ than its continuous counterpart $L^p(\T_1\times\R, L^p(\T_{2\pi}))$ for $p\in[1,\infty]$. For convenience, we introduce the discrete Bloch-Fourier transform
$\caB_d:=\caF_d\caB$; that is,
\beq
\label{e:dbft}
\begin{matrix}
\caB_d :& L^2(\R^2) &\longmapsto& L^2(\T_1 \times \R, \ell^2)\\
& v & \longrightarrow & \caB_d v(\bnu)  =(\caF_d\caB v)(\bnu)= \{ \widehat{v}(\nu_1 + j, \nu_2)\}_{j\in\Z},
\end{matrix}
\eeq
and the discrete version of $\widehat{\caL_p}$, denoted as 
\[
\widehat{\caL_d}=\int_{\T_1\times\R}\widehat{\caL_d}(\bnu)\rmd \bnu, \text{ where }
\widehat{\caL_d}(\bnu):=\caF_d\circ \widehat{\caL_p}(\bnu)\circ\caF_d^{-1};\]
that is,
\beq
\label{e:d-lin}
\begin{matrix}
\widehat{\caL_d}(\bnu) :&  w^{4,p} &\longrightarrow & \ell^p\\
&\underline{v}=\{v_j\}_{j\in \Z} &\longmapsto & \{\mu_j(\bnu;\kappa)v_j\}_{j\in\Z}+(\veps^2-\kappa^2)\underline{v} - 3\caF_d(u_p^2)*\underline{v},
\end{matrix}
\eeq
where $\mu_j(\bnu;\kappa):=-(1 - (1 + \kappa)(j + \nu_1)^2 - \nu_2^2)^2+\kappa^2$.
Introducing the notation 
\[
\underline{V}(\bnu):=\caB_d v=\{\widehat{v}(\nu_1+j,\nu_2)\}_{j\in\Z},
\]
the SHE with respect to the perturbation in the discrete Bloch-Fourier space takes the form
\beq
\label{e:SHE-dBF}
\underline{V}_t=\widehat{\caL_d}\underline{V}-3\underline{\widehat{u_p}}*\underline{V}^{*2}-\underline{V}^{*3},
\eeq
where the convolution in $\underline{V}^{*m}$ is with respect to $\bnu\in\T_1\times \R$ and $j\in\Z$, and the convolution with $\underline{\widehat{u_p}}$ is with respect to $j\in\Z$.
Before we introduce the mode filter decomposition, we first prove that the spectral properties of $\widehat{\caL_d}(\bnu)$ are independent of the choice of $p\in[1,\infty]$.
\begin{proposition} 
\label{p:sectorial}
For any $\bnu\in\T_1\times \R$ and $p\in[1,\infty]$, the closed operator $\widehat{\caL_d}(\bnu) : w^{4,p} \to \ell^p$ is sectorial with compact resolvents. More specifically, the sectoriality of $\widehat{\caL_d}(\bnu)$ is independent of the choice of $\bnu$ and $p$; that is, there exist $C>0$, $\omega\in(\pi/2, \pi)$ and $\lambda_0\in\R$, independent of $\bnu$ and $p$, such that 
\[
\opnorm{(\widehat{\caL_d}(\bnu)-\lambda)^{-1}}_{\ell^p}\leq \frac{C}{|\lambda-\lambda_0|}, \text{ for any }\lambda\in S(\lambda_0,\omega):=\left\{\lambda\in\C\;\middle|\;|\arg(\lambda-\lambda_0)|<\omega, \lambda\neq\lambda_0 \right\}.
\]
Moreover, the spectrum of $\widehat{\caL_d}(\bnu)$ is independent of the choice of the underlying space $\ell^p$ and thus denoted as $\sigma(\widehat{\caL_d}(\bnu))$, consisting only of isolated eigenvalues with finite multiplicities.
\end{proposition}
\begin{proof}
  See Appendix \ref{a:4}.
\end{proof}

\paragraph{Mode filter decomposition} 
We introduce the notation $\mathbb{X}^p := L^p(T_1 \times \mathbb{R}, \ell^p)$ and define an even smooth cut-off function  $\chi :\T_1\times\R \to [0,\infty)$ as
\beq
\label{e:cutoff}
  \chi(\bnu) = \begin{cases}
    1, & \abs{\bnu} \leq 1, \\
    0, & \abs{\bnu} \geq 2, \\
  \end{cases} 
\eeq
 as well as its rescaled version $\chi_\epsilon(\bnu):=\chi(\frac{\bnu}{\epsilon})$ for any $\epsilon>0$. We recall from Proposition \ref{p:spectral} that the eigenpair at $\bnu=0$ admits an analytic continuation for $|\bnu|<r_0$, and introduce the pseudo-eigenfunction 
\beq
\label{e:pseudo}
e_c(\bnu):=(1-\chi_{r_1}(\bnu))e_0+\chi_{r_1}(\bnu)e(\bnu),
\eeq
where $r_1:=r_0/4$,
and the projection mapping
\begin{equation}
    \begin{matrix}
      P: & \mathbb{X}^p & \longrightarrow & \mathbb{X}^p\\
      & \underline{V}&\longmapsto & \|\underline{\widehat{e_c}}\|_{\ell^2}^{-2}\llangle \underline{V},\widehat{\underline{e_c}}  \rrangle\underline{\widehat{e_c}}.
    \end{matrix}
\end{equation}
We readily see that $P$ is a bounded projection but not commutative with $\widehat{\caL_d}$; that is
\[
P^2=P, \qquad P\widehat{\caL_d}\neq\widehat{\caL_d}P.
\]
Introducing $Q:=\id-P$, $\mathbb{X}_c^p := \rg(P)$ and $\mathbb{X}_s^p :=\rg(\id - P)$, we note the subspaces $\mathbb{X}_c^p$ and $\mathbb{X}_s^p$ are closed in $\mathbb{X}$, and 
\[
\mathbb{X}^p=\mathbb{X}_c^p\bigoplus\mathbb{X}_s^p,
\]
in the sense that $\mathbb{X}^p = \mathbb{X}_c^p + \mathbb{X}_s^p$ and $\mathbb{X}_c^p \cap \mathbb{X}_s^p= \{0\}$.
Introducing the neutral and stable modes respectively,
\beq\label{e:aV-def}
a(t,\bnu):=K\llangle \underline{V}(t,\bnu),\widehat{\underline{e_c}}  \rrangle,\qquad \underline{V_s}(t,\bnu) := Q\underline{V}(t,\bnu),
\eeq
where $K(\bnu):=\|\underline{\widehat{e_c}}(\bnu)\|_{\ell^2}^{-2}=\|e_c(\bnu)\|_{L^2(\T_{2\pi})}^{-2}$ and the dependence of $K$ on $\bnu$ is typically suppressed for convenience. 
we apply the mode filter decomposition
\beq \label{e:Tmf}
T_{mf}\underline{V}:=\begin{pmatrix}
    a\\ \underline{V_s}
\end{pmatrix},
\eeq
which is an isomorphism from $\mathbb{X}^p$ to $L^p(\T_1\times \R^1)\times \mathbb{X}_s^p$,
to the SHE and rewrite the SHE in terms of the mode-filter coordinates $W:=(a,\underline{V_s})^T$; that is,
\beq
\label{e:sys}
W_t=L_{mf}W+N_{mf}(W),
\eeq
where 
\[
L_{mf}:= 
\begin{pmatrix} 
  L_{11}  &  L_{12}   \\
  L_{21}        &  L_{22}
\end{pmatrix}
, \qquad
N_{mf}(W):=
\begin{pmatrix} 
  N_c(W) \\
  N_s(W)
\end{pmatrix},
\]
with
\begin{align*}
L_{11}a&:=K\langle e_c,\caL_p e_c  \rangle a,\\
L_{12}\underline{V_s}&:=K\llangle \underline{V_s}, \widehat{\caL_d}\underline{\widehat{e_c}} \rrangle,\\
L_{21}a&:=\left(\widehat{\caL_d}\underline{\widehat{e_c}}- \underline{\widehat{e_c}}L_{11}\right)a,\\
L_{22}\underline{V_s}&:=\left(\widehat{\caL_d}-\underline{\widehat{e_c}}L_{12} \right) \underline{V_s},\\
N_c &:= -K\llangle 3 \underline{\widehat{u_p}}*(a\underline{\widehat{e_c}}  + \vsb)^{*2}+ (a\underline{\widehat{e_c}}  + \vsb)^{*3},\underline{\widehat{e_c}}  \rrangle, \\
N_s &:= -3\underline{\widehat{u_p}}*(a\underline{\widehat{e_c}} + \underline{V_s})^{*2} 
       - (a\underline{\widehat{e_c}} + \underline{V_s})^{*3} - N_{c} \underline{\widehat{e_c}} .
\end{align*}

\subsection{Linear Semigroup Estimates}
\label{ss:2.2}
We for now restrict ourselves to the initial value problem of the linearized flow of \eqref{e:sys}; that is,
\[
\begin{cases}
    W_t=L_{mf}W,\\
    W(0)=W_0=(a_0,\vsb_0)^T,
\end{cases}
\]
whose solution takes the form
\beq
\label{e:linsol}
W(t)=\rme^{L_{mf}t}W_0.
\eeq
Introducing the notation 
\[
M:=\rme^{L_{mf}t}=\begin{pmatrix}
    M_{11} & M_{12} \\
    M_{21} & M_{22}
\end{pmatrix},
\]
we study the temporal decay estimates of this semigroup and its physical derivatives on general $L^p$ spaces. Our analysis is split into two subcases; that is, when $\bnu$ is close to zero and when $\bnu$ is away from zero. More specifically, we rewrite $M$ as follows.
\[
M=\chi_{\frac{r_1}{2}}M+(1-\chi_{\frac{r_1}{2}})M.
\]
For  $|\bnu| \leq r_1$, we denote $ L_s(\bnu):=\widehat{\caL_d}(\bnu)\mid_{(\underline{e}(\bnu))^\perp}$ and have $L_{mf}$ reduce to the direct sum of the spectral projections along the central and stable directions; that is,
\[
L_{mf}(\bnu)=\begin{pmatrix}
    \lambda(\bnu) & 0\\ 0 & L_s(\bnu)
\end{pmatrix}
\]
and thus
\beq
\label{e:smM}
M=\begin{pmatrix}
    M_{11} & M_{12} \\
    M_{21} & M_{22}
\end{pmatrix}=\begin{pmatrix}
    \rme^{\lambda t} & 0 \\
    0 & \rme^{L_st}
\end{pmatrix}.
\eeq
Moreover, we have the following estimations for $\chi_{\frac{r_1}{2}}M$.
\begin{lemma} \label{lem:semigroup1}
For any $1\leq p\leq q\leq \infty$, there exists a positive constant C such that the cut-off analytic semigroup $\chi_{\frac{r_1}{2}} M_{11}(t)$ admits the estimates
\beq
\label{e:sg-neu}
\opnorm{\bnu^\alpha\chi_{\frac{r_1}{2}} M_{11}(t)}_{L^q \to L^p} \leq C(1+t)^{-[\frac{\alpha_1}{2}+\frac{\alpha_2}{4}+\frac{3}{4}(\frac{1}{p}-\frac{1}{q})]},
\eeq
where $\bnu^\alpha=\nu_1^{\alpha_1}\nu_2^{\alpha_2}$ with $\alpha=(\alpha_1,\alpha_2) \in\N^2$ and the space $L^p$ stands for $L^p(\T_1\times\R)$.
\end{lemma}
\begin{proof}
We recall from \eqref{eq:spec2} that, for $|\bnu|<r_0=4r_1$ and $\veps\in(0,\veps_1)$, 
\[
\lambda(\bnu) =-[4+\caO(\widetilde{a}^3)]\nu_1^2 -[1+\caO(\widetilde{a}^4)]\nu_2^4+\caO(\nu_1^4+\nu_2^6),
\]
from which we readily have that for $|\bnu|\leq r_1$ and $\veps\in(0,\veps_1)$, there exist constants $d_1$ and $d_2$, independent of the choice of $\bnu$ and $\veps$, such that
\[
\lambda(\bnu) \leq - d_1\nu_1^2 - d_2\nu_2^4.
\]
As a result, we conclude that, for any $1\leq p\leq q\leq \infty$,
\begin{align*}
\norm{\bnu^\alpha \chi_{\frac{r_1}{2}}\rme^{\lambda t} a}_{L^p} \; &\leq \; 
\norm[1]{\bnu^\alpha \chi_{\frac{r_1}{2}}\rme^{(-d_1\nu_1^2 - d_2\nu_2^4)t}}_{L^r}\norm{a}_{L^q} \; \leq \;
C(1+t)^{-[\frac{\alpha_1}{2}+\frac{\alpha_2}{4}+\frac{3}{4}(\frac{1}{p}-\frac{1}{q})]}\norm{a}_{L^q},
\end{align*}
where $\frac{1}{r}:=\frac{1}{p}-\frac{1}{q}$.
\end{proof}

\begin{lemma} \label{lem:semigroup2}
For any given $p\in[1,+\infty]$, there exists positive constants $C$ and $\lambda_1$, independent of the choice of $p$, such that the cut-off analytic semigroup $\chi_{\frac{r_1}{2}} M_{22}(t)$ admits the estimates
\beq\label{e:sg-ss}
\opnorm{\chi_{\frac{r_1}{2}} M_{22}(t)}_{L^p(\T_1\times\R, \mathbb{X}_s^p)} \leq C e^{-\lambda_1 t}.
\eeq
\end{lemma}
\begin{proof}
  For any $|\bnu|\leq r_1$, we have $M_{22}=\rme^{L_s t}$, which also takes the form
  \[
  M_{22}=\rme^{L_s t}=\frac{1}{2\pi\rmi}\int_\Gamma \rme^{\lambda t}(\lambda-L_s)^{-1}\rmd \lambda,
  \]
  where $\Gamma $ is a sectorial curve in the left half of the complex plane so that $\sigma(L_s)$ stays to the left of the $\Gamma$. Moreover, we choose $\Gamma$  independent of $\bnu$, and there exist $\lambda_1>0$ so that $\sup\{\re(\lambda)\mid \lambda\in\Gamma\}<-\lambda_1$. A proof similar to Proposition \ref{p:sectorial}, thus omitted, shows that there exists $C>0$, independent of $\bnu$ and $p$, so that, for any $\lambda\in\Gamma$, 
  \[
  \opnorm{(\lambda-L_s)^{-1}}_{\ell^p}\leq \frac{C}{|\lambda-\lambda_1|},
  \]
  which concludes the proof.
\end{proof}

For $\bnu$ away from zero, that is $|\bnu|>r_1/2$ , we have the following estimations for $(1-\chi_{\frac{r_1}{2}})M$. 
\begin{lemma} \label{lem:semigroup3}
  For any given $p\in[1,+\infty]$, there exists positive constants C and $\lambda_2$, independent of the choice of $p$ such that the analytic semigroup $(1-\chi_{\frac{r_1}{2}})M$ admits the estimates
  \beq\label{e:sg-sta}
  \opnorm{\bnu^\alpha(1-\chi_{\frac{r_1}{2}})M_{ij}}_{L^p \to L^p} \leq C t^{-\frac{\alpha_2}{4}} \rme^{-\lambda_2 t},  
  \eeq
for $i,j=1,2$ and $\alpha=(\alpha_1,\alpha_2)\in\N^2$.
\end{lemma}
\begin{proof}
The operator $L_{mf}(\bnu)$ is conjugate with $\widehat{\caL_d}(\bnu)$; that is,
$$L_{mf}(\bnu)=T_{mf}(\bnu)\widehat{\caL_d}(\bnu)T_{mf}^{-1}(\bnu),$$ 
and thus 
$$M(t)=\rme^{L_{mf}(\bnu)t}= T_{mf}(\bnu)\rme^{\widehat{\caL_d}(\bnu)t} T_{mf}^{-1}(\bnu).$$ 
Recalling that the maps $T_{mf}$ defined in \eqref{e:Tmf} and its inverse are uniformly bounded with respect to $\bnu$ and $p$, we are left to prove that 
\begin{equation}
\label{e:sg-far}
 \opnorm{\bnu^\alpha(1-\chi_{\frac{r_1}{2}})\rme^{\widehat{\caL_d}t}}_{L^p(\T_1\times\R,\ell^p)} \leq C t^{-\frac{\alpha_2}{4}} \rme^{-\lambda_2 t}.
\end{equation} 
To prove the inequality \eqref{e:sg-far}, we first infer from Proposition \ref{p:sectorial} that for $|\bnu|>\frac{r_1}{2}$  the operator $\widehat{\caL_d}(\bnu)$ is sectorial with 
\beq\label{e:Ld-far}
\sigma(\widehat{\caL_d}(\bnu))\subset (-\infty, -\widetilde{\lambda}_2),
\eeq
for some $\widetilde{\lambda}_2> 0$, indepedent of $p$ and $\bnu$. As a result, a proof similar to the one of Lemma \ref{lem:semigroup2}, thus omitted, shows that, for any $p\in[1,\infty]$ and $|\bnu|>\frac{r_1}{2}$, there exists $C>0$, independent of $p$ and $\bnu$,
\beq
\label{e:SG-Ld-far}
\opnorm{\rme^{\widehat{\caL_d}(\bnu)t}}_{\ell^p} \leq C e^{-\widetilde{\lambda}_2 t}.
\eeq
Moreover, we have improved spectral estimates of $\widehat{\caL_d}(\bnu)$ to absorb $\bnu^\alpha$ for $|\bnu|\gg1$. More specifically, we recall that $\sigma(\widehat{\caL_d}(\bnu))=\sigma(\widehat{\caL_p}(\bnu))$ and for any $\lambda\in \sigma(\widehat{\caL_d}(\bnu))$, there exists $k\in\Z$ such that 
\[
\lambda=-[\nu_2^2+(k+\nu_1^2)^2-1]^2+\caO(\veps^2);
\]
see \eqref{e:hot-est} and its subsequent discussion for details. As a result, we readily see that, for sufficiently large $R\gg1$ and any $|\bnu|>R$, there is a sharper spectral estimate than \eqref{e:Ld-far}; that is,
\[
\sigma(\widehat{\caL_d}(\bnu))\subset (-\infty, -\frac{1}{2}(\nu_2^4+1)),
\]
where we note that $\frac{1}{2}$ can be replaced with any number in $(0,1)$ by adjusting the size of $R$. Again, a proof similar to the one of Lemma \ref{lem:semigroup2}, thus omitted, shows that, for any $p\in[1,\infty]$ and $|\bnu|>R$, there exists $C>0$, independent of $p$ and $\bnu$,
\beq
\label{e:SG-Ld-farfar}
\opnorm{\bnu^\alpha\rme^{\widehat{\caL_d}(\bnu)t}}_{\ell^p } \leq C|\bnu|^\alpha \rme^{- \frac{1}{2}(\nu_2^4+1)t}.
\eeq
We now show that the estimate \eqref{e:sg-far} is true by exploiting the estimates \eqref{e:SG-Ld-far} and \eqref{e:SG-Ld-farfar}; that is,
\begin{equation}
    \begin{aligned}
        \norm{\bnu^\alpha(1-\chi_{\frac{r_1}{2}})\rme^{\widehat{\caL_d}t}\underline{V}}_{L^p(\T_1\times\R,\ell^p)}^p\leq & \int_{\frac{r_1}{2}\leq|\bnu|\leq R}\left(|\bnu|^\alpha \norm{\rme^{\widehat{\caL_d}(\bnu)t}\underline{V}}_{\ell^p}\right)^p\rmd \bnu+ \int_{|\bnu|> R}\left( \norm{\bnu^\alpha\rme^{\widehat{\caL_d}(\bnu)t}\underline{V}}_{\ell^p}\right)^p\rmd \bnu\\
        \overset{\eqref{e:SG-Ld-far}}{\leq} & C\int_{\frac{r_1}{2}\leq|\bnu|\leq R}\left(|\bnu|^\alpha \rme^{-\widetilde{\lambda}_2t}\norm{\underline{V}}_{\ell^p}\right)^p\rmd \bnu+ \int_{|\bnu|> R}\left( \norm{\bnu^\alpha\rme^{\widehat{\caL_d}(\bnu)t}\underline{V}}_{\ell^p}\right)^p\rmd \bnu\\
        \overset{\eqref{e:SG-Ld-farfar}}{\leq} & C\left[\rme^{-\widetilde{\lambda}_2pt}\|\underline{V}\|_{L^p(\T_1\times\R,\ell^p)}^p+ \int_{|\bnu|> R}\left( |\bnu|^\alpha\rme^{- \frac{1}{2}(\nu_2^4+1)t}\norm{\underline{V}}_{\ell^p}\right)^p\rmd \bnu\right]\\
        \leq& C\left(\rme^{-\widetilde{\lambda}_2pt}+t^{-\frac{\alpha_2p}{4}}\rme^{-\frac{p}{2}t} \right)\|\underline{V}\|_{L^p(\T_1\times\R,\ell^p)}^p\\
        \leq& Ct^{-\frac{\alpha_2p}{4}}\rme^{-\lambda_2 pt} \|\underline{V}\|_{L^p(\T_1\times\R,\ell^p)}^p,\\
    \end{aligned}
\end{equation}
where we take $\lambda_2:=\min\{\widetilde{\lambda}_2,\frac{1}{2}\}$ and conclude the proof.
\end{proof}

Taking advantage of Lemmas \ref{lem:semigroup1}, \ref{lem:semigroup2} and \ref{lem:semigroup3}, we summarize linear estimates results of the linear flow \eqref{e:linsol} in the proposition below.
\begin{proposition}\label{prop:linsem}
For any $1\leq p\leq q\leq \infty$, there exists positive constants $C$,$\lambda_2$  and $\lambda_3$ independent of the choice of $p$, such that the linear solution $W$ given in \eqref{e:linsol} admits the estimates
\begin{subequations}
\label{e:linest}
    \begin{align}
    \norm{W(t)}_{L^p}= &\norm{M(t)\begin{pmatrix}
    a_{0}\\ \underline{V_s}_{0}
    \end{pmatrix}}_{L^p} \leq C \begin{pmatrix}
   (1+ t)^{-\frac{3}{4}(\frac{1}{p}-\frac{1}{q})} & \rme^{-\lambda_{2}t}\\ \rme^{-\lambda_{2}t}& \rme^{-\lambda_{3}t}
    \end{pmatrix}\begin{pmatrix}
     \norm{a_{0}}_{L^p}+\norm{a_{0}}_{L^q}\\\norm{\underline{V_s}_{0}}_{L^p}
    \end{pmatrix}, \label{e:linest-1}\\
    \norm{\bnu^\alpha a(t)}_{L^p}\leq &C\left[(1+t)^{-[\frac{\alpha_1}{2}+\frac{\alpha_2}{4}+\frac{3}{4}(\frac{1}{p}-\frac{1}{q})]}\|a_0\|_{L^q}+t^{-\frac{\alpha_2}{4}}\rme^{-\lambda_2 t}\left( \norm{a_0}_{L^p}+\norm{\vsb_0}_{L^p} \right) \right]\label{e:linest-2}
\end{align}
\end{subequations}

\end{proposition}
\begin{proof}
We first recall from \eqref{e:smM} that $\chi_{\frac{r_1}{2}}M_{12}=0$ and have the following estimate
\[
{\allowdisplaybreaks
\begin{aligned}
    \norm{M(t)_{11}a_{0}+M(t)_{12}\underline{V_s}_{0}}_{L^p}=&\norm{(\chi_{\frac{r_1}{2}}M_{11}+(1-\chi_{\frac{r_1}{2}})M_{11})a_{0}+(\chi_{\frac{r_1}{2}}M_{12}+(1-\chi_{\frac{r_1}{2}})M_{12})\underline{V_s}_{0}}_{L^p},\\
    \leq & \norm{\chi_{\frac{r_1}{2}}M_{11}a_0}_{L^p}+\norm{(1-\chi_{\frac{r_1}{2}})M_{11}a_{0}}_{L^p}+\norm{(1-\chi_{\frac{r_1}{2}})M_{12}\underline{V_s}_{0}}_{L^p},\\
    \leq &C\left((1+t)^{-\frac{3}{4}(\frac{1}{p}-\frac{1}{q})}\norm{a_{0}}_{L^q}+ \rme^{-\lambda_{2}t}\norm{a_{0}}_{L^p} \right)+ C\rme^{-\lambda_{2}t}\norm{\underline{V_s}_{0}}_{L^p}, \\
    \leq &C(1+t)^{-\frac{3}{4}(\frac{1}{p}-\frac{1}{q})}\left(\norm{a_{0}}_{L^p}+\norm{a_{0}}_{L^q} \right)+ C\rme^{-\lambda_{2}t}\norm{\underline{V_s}_{0}}_{L^p}. \\
\end{aligned}
}
\]
Similarly, we recall from \eqref{e:smM} that $\chi_{\frac{r_1}{2}}M_{21}=0$ and have the following estimate 
\[
\begin{aligned}
    \norm{M(t)_{21}a_{0}+M(t)_{22}\underline{V_s}_{0}}_{L^p}=&\norm{(\chi_{\frac{r_1}{2}}M_{21}+(1-\chi_{\frac{r_1}{2}})M_{21})a_{0}+(\chi_{\frac{r_1}{2}}M_{22}+(1-\chi_{\frac{r_1}{2}})M_{22})\underline{V_s}_{0}}_{L^p}\\
    \leq & \norm{(1-\chi_{\frac{r_1}{2}})M_{21}a_{0}}_{L^p}+\norm{(1-\chi_{\frac{r_1}{2}})M_{22}\underline{V_s}_{0}}_{L^p}+\norm{(1-\chi_{\frac{r_1}{2}})M_{22}\underline{V_s}_{0}}_{L^p}\\
    \leq &C\rme^{-\lambda_{2}t}\norm{a_{0}}_{L^q}+ C\rme^{-\lambda_{1}t}\norm{\underline{V_s}_{0}}_{L^p} + C\rme^{-\lambda_{2}t}\norm{\underline{V_s}_{0}}_{L^p} \\
    \leq &C\rme^{-\lambda_{2}t}\norm{a_{0}}_{L^p}+ C\rme^{-\lambda_{3}t}\norm{\underline{V_s}_{0}}_{L^p} \\
\end{aligned}
\]
where $\lambda_{3}=\min\{\lambda_{1},\lambda_{2}\}.$ The conclusion of the proposition follows from the above estimations.
\end{proof}

\subsection{Global irrelevancy of nonlinear terms}
\label{ss:2.3}
 We produce a bound on the slowest decaying nonlinear term which allows us to show the global irrelevance of the nonlinear terms $N_{mf}(W)$ in the full nonlinear system \eqref{e:sys}. We used the term ``global irrelevance" to emphasize the fact that the perturbation $W$ preserves its leading order linear dynamics in the full nonlinear system \eqref{e:sys} while the nonlinear term $N_s(W)$ is ``locally relevant" in the nonlinear equation of  the stable entry $\underline{V_s}$.

To show the above global irrelevancy of $N_{mf}(W)$ and local relevancy of $N_s(W)$, we first single out the slowest decaying terms in $N_c(W)$ and $N_s(W)$ respectively. Recalling the fact from \eqref{e:aV-def} that 
\[
V_s(t,\bnu;\xi)=V(t,\bnu;\xi)-a(t,\bnu)e_c(\bnu;\xi),
\]
we expand the nonlinear terms $N_c$ and $N_s$ from their definitions given in Equation \eqref{e:sys}; that is,
\beq
\label{e:non-term}
\begin{aligned}
  N_c =
  &-3K\langle u_p(a e_c + V_s)^{*2},e_c \rangle 
       - K\langle (a e_c + V_s)^{*3},e_c \rangle, \\
  =
  &
  \overbrace{-3K\langle u_p(a e_c)^{*2},e_c \rangle}^{:=N_{c,1}(a,a)}
  \overbrace{- 6K\langle  u_p((a e_c)* V_s),e_c \rangle}^{:=N_{c,2}(a,\vsb)}
  \overbrace{-3K \langle u_p(V_s)^{*2},e_c \rangle}^{:=N_{c,3}(\vsb,\vsb)} \\
  &
  \overbrace{-K \langle (a e_c)^{*3}, e_c \rangle}^{:=N_{c,4}(a,a,a)}
    \overbrace{-3K\langle  (a e_c)^{*2} *V_s, e_c \rangle}^{:=N_{c,5}(a,a,\vsb)}
    \overbrace{-3K\langle (a e_c) * (V_s)^{*2}, e_c \rangle}^{:=N_{c,6}(a,\vsb,\vsb)}
    \overbrace{-K\langle (V_s)^{* 3}, e_c \rangle}^{:=N_{c,7}(\vsb,\vsb,\vsb)},
\end{aligned}
\eeq
\beq
\label{e:non-term2}
\begin{aligned}
      N_s =& -3\underline{\widehat{u_p}}*(a\eb + \underline{V_s})^{*2} 
       - (a\eb + \underline{V_s})^{*3} - N_{c} \;\eb,\\
     : = & N_{s,1}(a,a) + N_{s,2}(a,\vs) + N_{s,3}(\vs,\vs) +\\
    &N_{s,4}(a,a,a) + N_{s,5}(a,a,\vs) + N_{s,6}(a,\vs,\vs) + N_{s,7}(\vs,\vs,\vs),
\end{aligned}
\eeq
where each $N_{c,j}$ is a multilinear operator, and we define $N_{s,j}$ in a similar fashion and omit the details.

We now give intuitions on why the nonlinear term $N_{mf}(W)$ is globally irrelevant. We start with showing that $N_s(W)$ is locally relevant in the nonlinear equation of $\underline{V_s}$. We note that the linear flow of the stable component $\vsb$ decays exponentially and the leading order nonlinear term is $N_{s,1}(a,a)$. As a result, we have that in the nonlinear flow \eqref{e:sys}, the $L^1(\T_1\times\R, \ell^1)$ norm of $\vsb(t)$ has the same temporal decay as the $L^1(\T_1\times\R)$ norm of $a^{*2}$ as $t$ goes to $+\infty$; that is,
\[
\norm{\vsb(t,\cdot)}_{L^1(\T_1\times\R, \ell^1)}\sim \norm{a^{*2}}_{L^1(\T_1\times\R)}\sim t^{-\frac{3}{2}}.
\]
On the other hand, the temporal decay rate of the $L^1(\T_1\times \R )$ norm of the linear terms in the neutral mode equation of $a$ in \eqref{e:sys} is $t^{-\frac{7}{4}}$ and thus any irrelevant nonlinear terms should have a better temporal decay rate than that. Noting that, in the sense of the linear dynamics, 
\[
\|a(t,\cdot)\|_{L^1(\T_1\times\R)}\sim t^{-\frac{3}{4}},
\]
we derive that all nonlinear terms in $N_c$ are irrelevant except for the first term $N_{c,1}(a,a)$, which is seemingly relevant, based on the rough estimate $\norm{N_{c,1}}_{L^1(\T_1\times\R)}\sim \norm{a^{*2}}_{L^1(\T_1\times\R)}\sim t^{-\frac{3}{2}}$. But a careful analysis below reveals a refined structure of $N_{c,1}$, providing extra spatial derivative in the $x_1$ direction and thus rendering extra $t^{-\frac{1}{2}}$ decay, which in turn shows that $N_{c,1}$ is in fact irrelevant as well. More specifically, we have 
\beq
\label{e:Nc1-bound}
\begin{aligned}
  N_{c,1} &=-3K \langle u_p(ae_c)^{*2}, e_c \rangle
  \\
  &=
  -3 K\int_{T_{2\pi}} u_p(ae_c)^{*2} \overline{e_c} \rmd\xi \\
  &=
  -3K\int_{T_{2\pi}} u_{p}(\xi)\overline{e_c}(\bnu ; \xi)
  \left(
    \int\limits_{\bnut \in T_1 \times \mathbb{R}}
    a(\bnu-\bnut)e_c(\bnu-\bnut;\xi)  
    a(\bnut)e_c(\bnut;\xi) d\bnut
  \right)
  \rmd\xi \\
  &=
  \int\limits_{\bnut \in T_1 \times \mathbb{R}}
  \overbrace{-3K\left(
    \int_{T_{2\pi}}
    u_{p}(\xi)
    \overline{e_c}(\bnu;\xi)
    e_c(\bnut;\xi)
    e_c(\bnu-\bnut;\xi) \rmd \xi
  \right)}^{:=k_1(\bnu,\bnut,\bnu-\bnut)}
  a(\bnu-\bnut) a(\bnut) \rmd\bnut ,
\end{aligned}
\eeq
where we note $N_{c,1}$ is a weighted convolution with kernel $k_1$ and we suppressed $t$-dependence of $a$ for conveniences. Leveraging the parities in the expansion \eqref{eq:spec} of the eigenfunction $e$, we have the following refined estimate of the kernel $k_1$.
\begin{lemma} \label{lem:k1bound}
There exist positive constants $M, C>0$, independent of the choice of $\bnu, \widetilde{\bnu}\in\T_1\times\R$, such that 
\begin{subequations}\label{e:k1-est}
    \begin{align}
        \abs{k_1(\bnu, \bnut, \bnut-\bnu)}& \leq M, \label{e:k1-ub}\\
        \abs{k_1(\bnu, \bnut, \bnut-\bnu)} &\leq C\left( \; \abs{\nu_1-\widetilde{\nu}_1}+\abs{\widetilde{\nu}_1}. \label{e:k1-at0}
    \right).
    \end{align}
\end{subequations}
\end{lemma}
\begin{proof}
Recalling from \eqref{e:pseudo} that 
\[
e_c(\bnu;\xi)=(1-\chi_{r_1}(\bnu))e_0(\xi)+\chi_{r_1}(\bnu)e(\bnu;\xi),
\]
and the fact that $k_1=\int_{T_{2\pi}}
    u_{p}(\xi)
    \overline{e_c}(\bnu;\xi)
    e_c(\bnut;\xi)
    e_c(\bnu-\bnut;\xi) \rmd \xi$, we readily conclude that the integrand of $k_1$ is uniformly bounded for $\xi\in\T_{2\pi}$, $\bnu,\widetilde{\bnu}\in\T_1\times\R$ and thus there exists $M>0$ such that $|k_1(\bnu, \bnut, \bnut-\bnu)|\leq M$. 
To show the second inequality, we exploit the expansion of $e(\bnu)=e_r(\bnu)+\rmi\nu_1e_i(\bnu)$ in \eqref{eq:spec} to rewrite $e_c$ in \eqref{e:pseudo}; that is,
\beq
e_c(\bnu)=\overbrace{(1-\chi_{r_1}(\bnu))e_0+\chi_{r_1}(\bnu)e_r(\bnu)}^{:=\widetilde{e_r}}+\rmi\nu_1\overbrace{\chi_{r_1}e_i}^{:=\widetilde{e_i}}.
\eeq
where the real-valued functions $e_r$ and $e_i$ are respectively odd and even in $\xi$ for $|\bnu|<r_1$. Since $e_{0}$ and $e_r$ are odd, it follows that $\widetilde{e_r}$ is odd. Also since $e_i$ is even, so is $\widetilde{e_i}$. Noting that $u_p$ is even in $\xi$ and the integrand $u_p(\xi) \overline{e_c}(\bnu;\xi) e_c(\bnut;\xi)
  e_c(\bnut-\bnu;\xi)$ is $2\pi$-periodic in $\xi$, we readily see that the odd part vanish under the integration  on ${T_{2\pi}}$, yielding
  \begin{align*}
    k_1(\bnu, \bnut, \bnut-\bnu) 
    =& 
    \int_{T_{2\pi}} \rmi \nu_1 
    \left[3Ku_p \widetilde{e_i}(\bnu;\xi)\widetilde{e_r}(\bnut;\xi)\widetilde{e_r}(\bnu-\bnut;\xi)\right]
    \rmd\xi + \\
    &
    \int_{T_{2\pi}} \rmi\widetilde{\nu}_1 
    \left[-3Ku_p \widetilde{e_r}(\bnu;\xi)\widetilde{e_i}(\bnut;\xi)\widetilde{e_r}(\bnu-\bnut;\xi)\right]
    \rmd\xi + \\
    &
    \int_{T_{2\pi}} \rmi(\nu_1-\widetilde{\nu}_1) 
    \left[-3K u_p \widetilde{e_r}(\bnu;\xi) \widetilde{e_r}(\bnut;\xi)\widetilde{e_i}(\bnu - \bnut;\xi)\right]
    \rmd\xi + \\
    &
    \int_{T_{2\pi}} \rmi  (\nu_1-\widetilde{\nu}_1) 
    \left[ -3K\nu_1 \widetilde{\nu}_1u_p \widetilde{e_i}(\bnu;\xi) \widetilde{e_i}(\bnut;\xi)\widetilde{e_i}(\bnu - \bnut;\xi)\right]
    \rmd\xi,
  \end{align*}
  where all the four terms in the brackets are uniformly bounded in $\xi\in\T_{2\pi}$ and $\bnu, \widetilde{\bnu}\in\T_1\times\R$, thanks to the cutoff function $\xi_{r_1}$ in $\widetilde{e_i}$. As a result, there exists a positive constant $C$ independent of $\bnu, \bnut$ such that
  \[
    \abs{k_1(\bnu, \bnut, \bnut-\bnu)} \leq
    C
    \left(\;  \abs{\nu_1-\widetilde{\nu}_1}+\abs{\widetilde{\nu}_1}
    \right).
  \]
\end{proof}

Recalling that $\displaystyle \|a(t,\cdot)\|_{L^1(\T_1\times\R)}\sim t^{-\frac{3}{4}}$ and combining \eqref{e:Nc1-bound} and \eqref{e:k1-at0}, we have
\[
\norm{N_{c,1}}_{L^1(\T_1\times\R)}\sim \norm{a*(\nu_1 a)}_{L^1(\T_1\times\R)}\sim t^{-2},
\]
which decays faster than the linear terms and thus irrelevant.

\br
\label{r:gen}
Such a cancellation should be generically true for any pattern forming system posted on an unbounded spatial domain satisfies the following conditions:
\begin{itemize}
\item It admits translational symmetry (and rotational symmetry if the spatial domain is multi-dimensional) and no other hidden symmetry;
\item It gives rise to spatial periodic patterns,
\end{itemize}
In particular, similar arguments with different decay rates can be derived to show the global nonlinear irrelevancy in the case of the roll patterns $u_p(kx;k)$ with $k\in(k_z,k^+)$ of the 2D Swift-Hohenberg equatioin, rendering an alternative perspective besides the work \cite{uecker_1999}. Meanwhile, it seems that such an argument is only able to show that the leading order nonlinear term is critical for the 1D case, and thus requires more work to reproduce the classical result in \cite{schneider_1996}.
\er

\section{Proof of the main theorem}
\label{s:3}
We now give a proof of Theorem \ref{thm:main} via a fixed point argument on the variation of constant formula. More specifically, the solution to \eqref{e:sys} with the initial condition $W_0:=(a_0,\vsb_0)^T$ satisfies the variation of constant formula,
\beq
\label{eq:sshe}
  W(t) = \overbrace{\overbrace{e^{L_{mf}t}W_0}^{:=\caT_1(W_0)} + \overbrace{\int_0^t e^{L_{mf}(t-s)}N_{mf}(W(s)) \rmd s}^{:=\caT_2(W)}}^{:=\mathcal{T}(W)}.
\eeq
Our task is to show that $\caT$ is a well-defined contraction mapping on a bounded closed set in a proper Banach space, whose norm directly gives rise to the nonlinear weak diffusive decay in Theorem \ref{thm:main}.

\paragraph{The Space $\mathcal{H}$}
Based on our intuitive analysis on the irrelevancy of nonlinear terms in Section \ref{s:2}, we introduce the Banach space 
\beq
\label{e:H}
\H := \left\lbrace W(t,\bnu) =   
\begin{pmatrix}
  a(t,\bnu) \\
  \vsb(t,\bnu)
\end{pmatrix}
\Bigg|\norm{W}_\H < +\infty \right\rbrace,
\eeq
in which we introduction the following norms
\begin{align} 
  \norm{W}_{\H} \;
  &:= 
  \norm{a}_{\H_c} +\
  \norm[1]{\vsb}_{\H_s},  \label{eq:norm}\\
  \norm{a}_{\H_c} \;
  &:= 
  \sup_{t \geq 0} (1+t)^{3/4}\norm{a(t,\cdot)}_1 \; +\
  \sup_{t \geq 0} \norm{a(t,\cdot)}_\infty +\
  \sup_{t \geq 0} (1+t)^{5/4}\norm{\nu_1a(t,\cdot)}_1, \\
  \norm[1]{\vsb}_{\H_s}
  &:= 
  \sup_{t \geq 0} (1+t)^{3/2}\norm[1]{\vsb(t,\cdot)}_1 +\ 
  \sup_{t \geq 0} \norm[1]{\vsb(t,\cdot)}_\infty,
\end{align} 
where $\norm{a(t,\cdot)}_p := \norm{a(t, \cdot)}_{L^p(\T_1\times\R)}$ and $\norm{\vsb(t,\cdot)}_p:=\norm{\vsb(t,\cdot)}_{L^p(\T_1\times\R,\ell^p)}$. 

\br
In order to close the nonlinear argument, the appearance of $\nu_1 a(t,\cdot)$ term in $N_{c,1}$ is coped with by the introduction of  the term $\sup_{t \geq 0} (1+t)^{5/4}\norm{\nu_1a(t,\cdot)}_1$ in the norm.
\er

\paragraph{Linear estimates of $\caT_1$ in $\H$}
We first derive an upper bound of $\caT_1(W_0)$ in $\H$.
\begin{proposition}\label{thm:linearbound}
    There exist a positive constant $C$ such that
\beq\label{e:T1-est}
\norm{\caT_{1}(W_0)}_{\H}= \norm{M(t)W_0}_{\H} \leq C
    \left(
      \norm{W_0}_1 + \norm{W_0}_\infty
    \right).
\eeq
where $\norm{W_0}_1:=\norm{a_0}_1+\norm{\vsb_0}_1$ and $\norm{W_0}_\infty:=\norm{a_0}_\infty+\norm{\vsb_0}_\infty$.
\end{proposition} 
\begin{proof}
 By the definition of the $\H$-norm and the notation of the semigroup $M(t)$, we have
\[
\begin{aligned}
    \norm{\caT_1(W_0)}_{\H}=&\norm{M(t)\begin{pmatrix}
    a_{0}\\ \underline{V_s}_{0}\end{pmatrix}}_{\H},\\
    =&\norm{M(t)_{11}a_{0}+M(t)_{12}\underline{V_s}_{0}}_{\H_c}+\;\norm{M(t)_{21}a_{0}+M(t)_{22}\underline{V_s}_{0}}_{\H_s},\\
     =& \overbrace{\sup_{t \geq 0} (1+t)^{3/4}\norm{M(t)_{11}a_{0}+M(t)_{12}\underline{V_s}_{0}}_1}^{:= \Rmnum{1}} \; +\
 \overbrace{\sup_{t \geq 0} \norm{M(t)_{11}a_{0}+M(t)_{12}\underline{V_s}_{0}}_\infty}^{:=\Rmnum{2}} +  \\
  & \ \overbrace{\sup_{t \geq 0} (1+t)^{5/4}\norm{\nu_1\left(M(t)_{11}a_{0}+M(t)_{12}\underline{V_s}_{0}\right)}_1}^{:=\Rmnum{3}} +\overbrace{\sup_{t \geq 0} (1+t)^{3/2}\norm[1]{M(t)_{21}a_{0}+M(t)_{22}\underline{V_s}_{0}}_1}^{:=\Rmnum{4}}  +\\
  &\ \overbrace{\sup_{t \geq 0} \norm[1]{M(t)_{21}a_{0}+M(t)_{22}\underline{V_s}_{0}}_\infty}^{:=\Rmnum{5}}. 
\end{aligned}
\]
Taking advantage of Proposition \ref{prop:linsem}, we derive upper bounds of the terms $\Rmnum{1}-\Rmnum{5}$ respectively. 
\begin{enumerate}
\item We exploit the estimates of $M_{11}$ and $M_{12}$ in \eqref{e:linest-1} with $p=1$ and $q=\infty$, yielding
\[
\begin{aligned}
        \Rmnum{1} &\leq \sup_{t \geq 0} (1+t)^{-\frac{3}{4}}\left((1+t)^{\frac{3}{4}}\left(\norm{a_{0}}_{1}+\norm{a_{0}}_{\infty} \right)+ \rme^{-\lambda_{2}t}\norm{\underline{V_s}_{0}}_{1}  \right)\\
        & \leq C \left( \norm{a_0}_1 + \norm{a_0}_\infty + 
      \norm{\vsb_0}_1\right).
\end{aligned}
\]
\item We exploit the estimates of $M_{11}$ and $M_{12}$ in \eqref{e:linest-1} with $p=q=\infty$, yielding
\[
\begin{aligned}
        \Rmnum{2} &\leq \sup_{t \geq 0} \left(\norm{a_{0}}_{\infty} + \rme^{-\lambda_{2}t}\norm{\underline{V_s}_{0}}_{\infty}  \right)  \leq C \left( \norm{a_0}_\infty + 
      \norm{\vsb_0}_{\infty}\right).
\end{aligned}
\]
\item We exploit the estimate \eqref{e:linest-2} with $p=1$, $q=\infty$ and $\alpha=(1,0)$, yielding
\[
\begin{aligned}
    \Rmnum{3}&= \sup_{t \geq 0} (1+t)^{\frac{5}{4}}\norm{\nu_1\left(M(t)_{11}a_{0}+M(t)_{12}\underline{V_s}_{0}\right)}_1\\
        &\leq C \sup_{t \geq 0} (1+t)^{-\frac{5}{4}}\left((1+t)^{\frac{5}{4}}\norm{a_{0}}_{\infty}+\rme^{-\lambda_{2}t}\norm{a_{0}}_{1} + \rme^{-\lambda_{2}t}\norm{\underline{V_s}_{0}}_{1}  \right) \\
        & \leq C \left(\norm{a_{0}}_{\infty}+ \norm{a_0}_1 + 
      \norm{\vsb_0}_{1}\right).
\end{aligned}
\]
\item We exploit the estimates of $M_{21}$ and $M_{22}$ in \eqref{e:linest-1} with $p=1$ and $p=\infty$ respectively, yielding
\[
\begin{aligned}
        \Rmnum{4} &\leq \sup_{t \geq 0}(1+t)^{\frac{3}{2}}\left(\rme^{-\lambda_{2}t}\norm{a_{0}}_{1} + \rme^{-\lambda_{3}t}\norm{\underline{V_s}_{0}}_{1}  \right) \leq C \left( \norm{a_0}_1 + 
      \norm{\vsb_0}_{1}\right),\\
      \Rmnum{5} &\leq \sup_{t \geq 0} \left(\rme^{-\lambda_{2}t}\norm{a_{0}}_{\infty} + \rme^{-\lambda_{3}t}\norm{\underline{V_s}_{0}}_{\infty}  \right)\leq C \left( \norm{a_0}_\infty + 
      \norm{\vsb_0}_{\infty}\right).
\end{aligned}
\]
\end{enumerate}
Combining the above estimates concludes the proof.
\end{proof}


\paragraph{Nonlinear estimates of $\caT_2$ in $\H$} We now show that $\caT_2(W)$ is bounded in $\H$. More specifically, we have the following proposition.
\begin{proposition}
    \label{prop:nonlinearbound}
   There exists $C>0$ such that, for any $W\in \H$,
  \beq\label{e:T2-est}
    \norm{\caT_2(W)}_\H 
    \leq
    C
    \left(
      \norm{W}_\H^2 + \norm{W}_\H^3
    \right).
  \eeq
  Moreover, $\caT_2$ is locally Lipschitz continuous in the sense that there exists $C>0$ such that, for any $W_1, W_2\in \H$,
  \beq\label{e:T2-lip}
  \norm{\caT_2(W_1)-\caT_2(W_2)}_\H 
    \leq
    C\|W_1-W_2\|_{\H}
    \left(
      \norm{W_1}_\H + \norm{W_2}_\H + \norm{W_1}_\H^2+ \norm{W_2}_\H^2
    \right).
  \eeq
\end{proposition} 

\begin{remark}
    We note that the estimate \eqref{e:T2-est} is only a special case of the Lipschitz continuity estimate \eqref{e:T2-lip} when $W_1=W$ and $W_2=0$. The reason why we keep the estimate \eqref{e:T2-est} is two-fold. Firstly, we need to use \eqref{e:T2-est} in our fixed point argument. Secondly, it is more natural to prove \eqref{e:T2-est} first and observe that the more general estimate \eqref{e:T2-lip} is a natural consequence of the special estimate \eqref{e:T2-est} by exploiting the fundamental theorem of calculus.
\end{remark}

\begin{proof}
    To prove the estimate \eqref{e:T2-est}, we first recall that 
    \[
    \caT_2(W)=\int_0^t e^{L_{mf}(t-s)}N_{mf}(W(s)) \rmd s=\int_0^t\begin{pmatrix}
        M_{11}(t-s) & M_{12}(t-s)\\ M_{21}(t-s) & M_{22}(t-s)
    \end{pmatrix}\begin{pmatrix}
        N_c(W(s))\\ N_s(W(s))
    \end{pmatrix}\rmd s,
    \]
    where we already have the estimates on the semigroup $M(t)$ in Lemma \ref{lem:semigroup1}-\ref{lem:semigroup3} and Proposition \ref{prop:linsem} and estimates on the nonlinear terms are yet to be given. We claim that the $L^1(\T_1\times\R)$-norm and $L^\infty(\T_1\times\R)$-norm of nonlinear terms $N_c$ and $N_s$ admits the following estimates
    \begin{subequations}
        \label{e:Ncs}
\begin{align}
  \norm{N_{c}(t,\cdot)}_1  \leq &C (1+t)^{-2}\left(\norm{W}_{\H}^2+\norm{W}_{\H}^3\right), \qquad &\norm{N_{c}(t,\cdot)}_{\infty}\leq C (1+t)^{-\frac{5}{4}}\left(\norm{W}_{\H}^2+\norm{W}_{\H}^3\right);\label{e:Nc}\\
  \norm{N_{s}(t,\cdot)}_1  \leq &C (1+t)^{-\frac{3}{2}}\left(\norm{W}_{\H}^2+\norm{W}_{\H}^3\right), \qquad &\norm{N_{s}(t,\cdot)}_{\infty}\leq C (1+t)^{-\frac{3}{4}}\left(\norm{W}_{\H}^2+\norm{W}_{\H}^3\right).\label{e:Ns}
\end{align}
    \end{subequations}
 To prove \eqref{e:Nc}, we recall the expansion $N_c =\sum_{j=1}^7 N_{c,j}$ in \eqref{e:non-term} and derive $L^1$ and $L^\infty$ estimates of $N_{c, j}$ respectively. Due to the hidden refined structure of $N_{c,1}$ as shown in \eqref{e:k1-at0}, we derive in details the estimates of $N_{c,1}$, from which the estimates of the rest $N_{c,j}$ terms follow naturally. More specifically, using the notations $N_{c,1}(W(t))$ and $N_{c,1}(t,\bnu)$ interchangeably, we have
\begin{equation}
    \label{e:Nc1}
    \begin{aligned}
        |N_{c,1}(t,\bnu)|=&\left|\int_{\T_1\times\R}k_1(\bnu,\bnut,\bnu-\bnut)a(t,\bnu-\bnut)a(t,\bnut)\rmd \bnut\right|\\
        \overset{\eqref{e:k1-at0}}{\leq}&C\left(\int_{\T_1\times\R}\left|a(t,\bnu-\bnut)\right|\left|\nu_1a(t,\bnut)\right|\rmd \bnut\right)\\
        =&C |a|*|\nu_1 a|,
    \end{aligned}
\end{equation}
which, via the Young's inequality for convolution and the definition of $\|\cdot\|_{\H_c}$, yields that
\beq
\label{e:Nc1-1}
\begin{aligned}
  \norm{N_{c,1}(t,\cdot)}_1  \leq &C\norm{a(t,\cdot)}_1\norm{\nu_1a(t,\cdot)}_1 
  \leq C  (1+t)^{-\frac{3}{4}}\norm{a}_{\H_c} (1+t)^{-\frac{5}{4}}\norm{a}_{\H_c}
  = C (1+t)^{-2}\norm{a}_{\H_c}^2; \\
  \norm{N_{c,1}(t,\cdot)}_{\infty}\leq & C
  \norm{a(t,\cdot)}_\infty\norm{\nu_1a(t,\cdot)}_1 
  \leq   C\norm{a}_{\H_c} (1+t)^{-\frac{5}{4}}\norm{a}_{\H_c}  = C (1+t)^{-\frac{5}{4}}\norm{a}_{\H_c}^2.
\end{aligned}
\eeq
For $N_{c,j}$, $2\leq j \leq 7$, recalling the definition of $N_{c,j}$ in \eqref{e:non-term}, similarly to $N_{c,1}$,
we can always rewrite $N_{c,j}$ in the form of weighted convolutions of $a$ and $V_s$ with resepct to $\bnu\in\T_1\times 
\R$ by taking integration with respect to $\xi\in\T_{2\pi}$ first. Noting that all the weight functions are uniformly bounded and that the product of $2\pi$-periodic functions in $\xi$ corresponds to the zero Fourier mode of a convolution in discrete $\ell^p$ spaces, we conclude that 
\begin{equation}
    \label{e:Ncj}
    \begin{aligned}
        |N_{c,j}(t,\bnu)|
        \leq& C |a|^{*m_j}*\norm{\vsb}_{\ell^1}^{*(n_j-k_j)}*\norm{\vsb}_{\ell^\infty}^{k_j},
    \end{aligned}
\end{equation}
where $k_j:=\min\{1, n_j\}$, the convolutions are with respect to $\bnu\in\T_1\times\R$ and the nonnegative integers $m_j$, $n_j$ satisfies that $m_j+n_j=2$ for $j=2,3$; and $m_j+n_j=3$ for $4\leq j\leq 7$. In combination with the Young's inequality for convolution and the definition of $\|\cdot\|_{\H_c}$, \eqref{e:Ncj} yields that
\beq
\label{e:Ncj-1}
\begin{aligned}
  \norm{N_{c,j}(t,\cdot)}_1  \leq &C\norm{a(t,\cdot)}_1^{m_j}\norm{\vsb(t,\cdot)}_1^{n_j} 
  \leq C  (1+t)^{-(\frac{3}{4}m_j+\frac{3}{2}n_j)}\norm{W}_{\H}^{m_j+n_j}; \\
  \norm{N_{c,j}(t,\cdot)}_{\infty}\leq & C
  \norm{a(t,\cdot)}_\infty^{\widetilde{k}_j}\norm{\vsb(t,\cdot)}_\infty^{1-\widetilde{k}_j}\norm{a(t,\cdot)}_1^{m_j-\widetilde{k}_j}\norm{\vsb(t,\cdot)}_{1}^{n_j+\widetilde{k}_j-1} \\
  \leq & C (1+t)^{-\frac{3}{4}(m_j-\widetilde{k}_j)-\frac{3}{2}(n_j+\widetilde{k}_j-1)}\norm{W}_{\H}^{m_j+n_j},
\end{aligned}
\eeq
where $\widetilde{k}_j:=\min\{1, m_j\}$. Noting that for $j>1$, 
\[
\frac{3}{4}m_j+\frac{3}{2}n_j\geq \frac{9}{4}, \qquad \frac{3}{4}(m_j-\widetilde{k}_j)+\frac{3}{2}(n_j+\widetilde{k}_j-1)\geq \frac{3}{2},
\]
we conclude from the estimates \eqref{e:Nc1-1} and \eqref{e:Ncj-1} that \eqref{e:Nc} is true.

The proof of \eqref{e:Ns} is similar to the one of \eqref{e:Nc}. We recall the expansion $N_s =\sum_{j=1}^7 N_{s,j}$ in \eqref{e:non-term2} and derive $L^1$ and $L^\infty$ estimates of $N_{s, j}$ respectively. Based on the above analysis of $N_c$, we readily see that, for any $1\leq j\leq 7$, $t>0$ and $\bnu\in\T_1\times\R$, 
\begin{equation}
    \label{e:Nsj}
    \begin{aligned}
        \norm{N_{s,j}(t,\bnu)}_{\ell^1}
        \leq& C |a|^{*m_j}*\norm{\vsb}_{\ell^1}^{*n_j},\\
        \norm{N_{s,j}(t,\bnu)}_{\ell^\infty}
        \leq& C |a|^{*m_j}*\norm{\vsb}_{\ell^1}^{*(n_j-k_j)}*\norm{\vsb}_{\ell^\infty}^{k_j},\\
    \end{aligned}
\end{equation}
which, via the Young's inequality for convolution, yields that
\beq
\label{e:Nsj-1}
\begin{aligned}
  \norm{N_{s,j}(t,\cdot)}_1  \leq &C\norm{a(t,\cdot)}_1^{m_j}\norm{\vsb(t,\cdot)}_1^{n_j} 
  \leq C  (1+t)^{-(\frac{3}{4}m_j+\frac{3}{2}n_j)}\norm{W}_{\H}^{m_j+n_j}; \\
  \norm{N_{s,j}(t,\cdot)}_{\infty}\leq & C
  \norm{a(t,\cdot)}_\infty^{\widetilde{k}_j}\norm{\vsb(t,\cdot)}_\infty^{1-\widetilde{k}_j}\norm{a(t,\cdot)}_1^{m_j-\widetilde{k}_j}\norm{\vsb(t,\cdot)}_{1}^{n_j+\widetilde{k}_j-1} \\
  \leq & C (1+t)^{-\frac{3}{4}(m_j-\widetilde{k}_j)-\frac{3}{2}(n_j+\widetilde{k}_j-1)}\norm{W}_{\H}^{m_j+n_j},
\end{aligned}
\eeq
which in turns leads directly to \eqref{e:Ns}.

We now prove \eqref{e:T2-est}. By the definition of the $\H$ norm, we have
{\allowdisplaybreaks
\begin{align}\label{e:T2-gen}
    \norm{\caT_2(W)}_{\H}=&\norm{\int_0^t\begin{pmatrix}
        M_{11}(t-s)N_c(W(s))+ M_{12}(t-s) N_s(W(s))\\ M_{21}(t-s) N_c(W(s)) +M_{22}(t-s) N_s(W(s)) \end{pmatrix} \rmd s}_{\H} \nonumber \\ 
    =&\norm{\int_0^t \left( M_{11}(t-s)N_c(W(s))+ M_{12}(t-s) N_s(W(s)) \right) \rmd s}_{\H_c} + \nonumber \\
    & \norm{\int_0^t \left( M_{21}(t-s) N_c(W(s)) + M_{22}(t-s) N_s(W(s)) \right)\rmd s}_{\H_s} \\
    \leq&\overbrace{\left(\norm{\int_0^tM_{11}(t-s)N_{c}(W(s))\rmd s}_{\H_c}+\norm{\int_0^tM_{21}(t-s)N_{c}(W(s))\rmd s}_{\H_s} \right) }^{:=\Rmnum{1}_{c}}+ \nonumber\\
    &\overbrace{\left(\norm{\int_0^tM_{12}(t-s)N_{s}(W(s))\rmd s}_{\H_c}+\norm{\int_0^tM_{22}(t-s)N_{s}(W(s))\rmd s}_{\H_s} \right) }^{:=\Rmnum{1}_{s}}. \nonumber
\end{align}
}

\underline{\textit{Estimate of $\Rmnum{1}_{c}$ }} We evaluate $\Rmnum{1}_{c}$ for small and large $\bnu$ respectively; that is,
\beq
\label{e:Ic}
\begin{aligned}
  \Rmnum{1}_{c}=&  \norm{\int_0^tM_{11}(t-s)N_{c}(W(s))\rmd s}_{\H_c}+\norm{\int_0^tM_{21}(t-s)N_{c}(W(s))\rmd s}_{\H_s}\\
  \leq&\overbrace{\norm{\int_0^t\chi_{\frac{r_1}{2}}M_{11}(t-s)N_{c}(W(s))\rmd s}_{\H_c}}^{:=\Rmnum{1}_{c,1}}+\overbrace{\norm{\int_0^t(1-\chi_{\frac{r_1}{2}})M_{11}(t-s)N_{c}(W(s))\rmd s}_{\H_c}}^{:=\Rmnum{1}_{c,2}}+\\
  &\overbrace{\norm{\int_0^t (1-\chi_{\frac{r_1}{2}})M_{21}(t-s)N_{c}(W(s))\rmd s}_{\H_s}}^{:=\Rmnum{1}_{c,3}},\\
\end{aligned}
\eeq
where we use the fact that $\chi_{\frac{r_1}{2}}M_{21}=0$. Moreover, recalling the definition of $\norm{\cdot}_{\H_c}$ and $\norm{\cdot}_{\H_s}$, we have
{\allowdisplaybreaks
\begin{align}\label{e:Icj}
    \Rmnum{1}_{c,1}\leq&\overbrace{\sup_{t \geq 0} (1+t)^{\frac{3}{4}}\int_0^t \norm{\chi_{\frac{r_1}{2}} M_{11}(t-s)N_{c}(W(s))}_1 \rmd s}^{:=A_{c,1}}+\overbrace{\sup_{t \geq 0} \int_0^t \norm{\chi_{\frac{r_1}{2}} M_{11}(t-s)N_{c}(W(s))}_\infty \rmd s}^{:=B_{c,1}}+ \nonumber\\
    &\overbrace{\sup_{t \geq 0}  (1+t)^{\frac{5}{4}}\int_0^t \norm{\nu_1\chi_{\frac{r_1}{2}} M_{11}(t-s)N_{c}(W(s))}_1 \rmd s}^{:=C_{c,1}}, \nonumber\\
    \Rmnum{1}_{c,2}\leq &\overbrace{\sup_{t \geq 0} (1+t)^{\frac{3}{4}}\int_0^t \norm{(1-\chi_{\frac{r_1}{2}}) M_{11}(t-s)N_{c}(W(s))}_1 \rmd s}^{:=A_{c,2}}+ \nonumber\\
    &\overbrace{\sup_{t \geq 0} \int_0^t \norm{(1-\chi_{\frac{r_1}{2}}) M_{11}(t-s)N_{c}(W(s))}_\infty \rmd s}^{:=B_{c,2}}+\\
    &\overbrace{\sup_{t \geq 0}  (1+t)^{\frac{5}{4}}\int_0^t \norm{\nu_1(1-\chi_{\frac{r_1}{2}}) M_{11}(t-s)N_{c}(W(s))}_1 \rmd s}^{:=C_{c,2}}, \nonumber\\
    \Rmnum{1}_{c,3}\leq &\overbrace{\sup_{t \geq 0}  (1+t)^{\frac{3}{2}} \int_0^t \norm{(1-\chi_{\frac{r_1}{2}})M_{21}(t-s)N_{c}(W(s))}_1 \rmd s}^{:=D_{c,3}}+ \nonumber\\
&\overbrace{\sup_{t \geq 0} \int_0^t \norm{(1-\chi_{\frac{r_1}{2}})M_{21}(t-s)N_{c}(W(s))}_\infty \rmd s}^{:=E_{c,3}}. \nonumber
\end{align}
}
In other words, we have 
\beq\label{e:Ic-int}
\begin{aligned}
    \Rmnum{1}_{c}\leq \sum_{j=1}^3 I_{c,j}\leq \sum_{j=1}^2\left(A_{c,j}+B_{c,j}+C_{c,j} \right)+ D_{c,3}+E_{c,3}.
\end{aligned}
\eeq
We are left to estimate all the terms in the right hand side of \eqref{e:Ic-int}. Taking advantage of the neutral mode estimate \eqref{e:sg-neu} and the estimate \eqref{e:Nc} of $N_c$, we have
{\allowdisplaybreaks
\begin{align}\label{e:ABCc1}
    A_{c,1}=&\sup_{t \geq 0} (1+t)^{\frac{3}{4}}\int_0^t \norm{\chi_{\frac{r_1}{2}} M_{11}(t-s)N_{c}(W(s))}_1 \rmd s \nonumber\\
 \leq&\sup_{t \geq 0} (1+t)^{\frac{3}{4}}
          \Bigg( \int_0^{t/2} \norm{\chi_{\frac{r_1}{2}} M_{11}(t-s)}_{L^\infty \to L^1}
          \norm{N_{c}(W(s))}_\infty
          \rmd s + \nonumber\\
          &\int_{t/2}^t \norm{\chi_{\frac{r_1}{2}} M_{11}(t-s)}_{L^1 \to L^1}
          \norm{N_{c}(W(s))}_1
          \rmd s \Bigg) \nonumber\\
        \overset{\eqref{e:sg-neu},\;\eqref{e:Nc}}{\leq}& C\left(\norm{W}_{\H}^2+\norm{W}_{\H}^3\right)\sup_{t \geq 0}(1+t)^{\frac{3}{4}} 
        \left(
          \int_0^{t/2} (1+t-s)^{-\frac{3}{4}}(1+s)^{-\frac{5}{4}} \rmd s +
          \int_{t/2}^t (1+s)^{-2} \rmd s 
        \right) \nonumber\\
        \leq & C \left(\norm{W}_{\H}^2+\norm{W}_{\H}^3\right); \nonumber\\
B_{c,1}=&\sup_{t \geq 0} \int_0^t \norm{\chi_{\frac{r_1}{2}} M_{11}(t-s)N_{c}(W(s))}_\infty \rmd s \nonumber\\
 \leq&\sup_{t \geq 0}\int_0^{t} \norm{\chi_{\frac{r_1}{2}} M_{11}(t-s)}_{L^\infty \to L^\infty}
          \norm{N_{c}(W(s))}_\infty
          \rmd s \nonumber\\
        \overset{\eqref{e:sg-neu},\;\eqref{e:Nc}}{\leq}& C\left(\norm{W}_{\H}^2+\norm{W}_{\H}^3\right)\sup_{t \geq 0} 
         \Bigg( \int_{0}^t (1+s)^{-5/4} \rmd s \Bigg)
         \\
        \leq & C \left(\norm{W}_{\H}^2+\norm{W}_{\H}^3\right);\nonumber\\
        C_{c,1}=&\sup_{t \geq 0}  (1+t)^{\frac{5}{4}}\int_0^t \norm{\nu_1\chi_{\frac{r_1}{2}} M_{11}(t-s)N_{c}(W(s))}_1 \rmd s \nonumber\\
 \leq&\sup_{t \geq 0} (1+t)^{\frac{5}{4}}\Bigg( \int_0^{t/2} \norm{\nu_1\chi_{\frac{r_1}{2}} M_{11}(t-s)}_{L^\infty \to L^1}
          \norm{N_{c}(W(s))}_\infty
          \rmd s + \nonumber\\
          &\int_{t/2}^t \norm{\nu_1\chi_{\frac{r_1}{2}} M_{11}(t-s)}_{L^1 \to L^1}
          \norm{N_{c}(W(s))}_1
          \rmd s \Bigg)\nonumber\\
        \overset{\eqref{e:sg-neu},\;\eqref{e:Nc}}{\leq}& C\left(\norm{W}_{\H}^2+\norm{W}_{\H}^3\right)\cdot\nonumber\\
        &\sup_{t \geq 0}(1+t)^{\frac{5}{4}} 
        \left(
          \int_0^{t/2} (1+t-s)^{-\frac{5}{4}}(1+s)^{-\frac{5}{4}} \rmd s +
          \int_{t/2}^t (1+t-s)^{-\frac{1}{2}}(1+s)^{-2} \rmd s 
        \right) \nonumber\\
        \leq & C \left(\norm{W}_{\H}^2+\norm{W}_{\H}^3\right),\nonumber
\end{align}
}
Similarly, taking advantage of the estimates \eqref{e:sg-sta} and \eqref{e:Nc}, we have
\beq
\label{e:ABCc2}
\begin{aligned}
   A_{c,2} =&\sup_{t \geq 0} (1+t)^{\frac{3}{4}} \int_0^t \norm{(1-\chi_{\frac{r_1}{2}})M_{11}(t-s)N_{c}(W(s))}_1 \rmd s\\
    \leq & C\sup_{t \geq 0} (1+t)^{\frac{3}{4}} \int_0^t \norm{(1-\chi_{\frac{r_1}{2}})M_{11}(t-s)}_1 \norm{N_{c}(W(s))}_1 \rmd s \\
   \overset{\eqref{e:sg-sta},\;\eqref{e:Nc}}{\leq} & C\left(\norm{W}_{\H}^2+\norm{W}_{\H}^3\right)\sup_{t \geq 0}(1+t)^{\frac{3}{4}}  \left(
          \int_0^{t} \rme^{-\lambda_2 (t-s)}(1+s)^{-2} \rmd s \right) \\
   \leq  & C\left(\norm{W}_{\H}^2+\norm{W}_{\H}^3\right)\sup_{t \geq 0}(1+t)^{\frac{3}{4}}  \left(
          \rme^{-\frac{\lambda_2t}{2}}\int_0^{t/2} (1+s)^{-2} \rmd s+(1+t/2)^{-2}\int_{t/2}^{t} \rme^{-\lambda_2 (t-s)} \rmd s \right) \\
   \leq &   C \left(\norm{W}_{\H}^2+\norm{W}_{\H}^3\right);\\
   B_{c,2}=&\sup_{t \geq 0}  \int_0^t \norm{(1-\chi_{\frac{r_1}{2}})M_{11}(t-s)N_{c}(W(s))}_\infty \rmd s\\
    \leq & C\sup_{t \geq 0} \int_0^t \norm{(1-\chi_{\frac{r_1}{2}}) M_{11}(t-s)}_{L^\infty \to L^\infty}\norm{N_{c}(W(s))}_\infty \rmd s \\
   \overset{\eqref{e:sg-sta},\;\eqref{e:Nc}}{\leq} & C\left(\norm{W}_{\H}^2+\norm{W}_{\H}^3\right)\sup_{t \geq 0} \left(  \int_0^{t} \rme^{-\lambda_2 (t-s)}(1+s)^{-\frac{5}{4}} \rmd s \right) \\
   \leq  & C\left(\norm{W}_{\H}^2+\norm{W}_{\H}^3\right)\sup_{t \geq 0} \left(
          \rme^{-\frac{\lambda_2t}{2}}\int_0^{t/2} (1+s)^{-\frac{5}{4}} \rmd s+(1+t/2)^{-\frac{5}{4}}\int_{t/2}^{t} \rme^{-\lambda_2 (t-s)} \rmd s \right) \\
   \leq &   C \left(\norm{W}_{\H}^2+\norm{W}_{\H}^3\right);\\
  C_{c,2}=& \sup_{t \geq 0}  (1+t)^{\frac{5}{4}} \int_0^t \norm{\nu_1(1-\chi_{\frac{r_1}{2}})M_{11}(t-s)N_{c}(W(s))}_1 \rmd s\\
    \leq & C\sup_{t \geq 0}  (1+t)^{\frac{5}{4}}\int_0^t \norm{\nu_1(1-\chi_{\frac{r_1}{2}}) M_{11}(t-s)}_{L^1 \to L^1}\norm{N_{c}(W(s))}_1 \rmd s \\
   \overset{\eqref{e:sg-sta},\;\eqref{e:Nc}}{\leq} & C\left(\norm{W}_{\H}^2+\norm{W}_{\H}^3\right)\sup_{t \geq 0} (1+t)^{\frac{5}{4}} \left(
          \int_0^{t} \rme^{-\lambda_2 (t-s)}(1+s)^{-2} \rmd s \right) \\
   \leq  & C\left(\norm{W}_{\H}^2+\norm{W}_{\H}^3\right)\sup_{t \geq 0} (1+t)^{\frac{5}{4}} \left(
          \rme^{-\frac{\lambda_2t}{2}}\int_0^{t/2} (1+s)^{-2} \rmd s+(1+t/2)^{-2}\int_{t/2}^{t} \rme^{-\lambda_2 (t-s)} \rmd s \right) \\       
   \leq &   C \left(\norm{W}_{\H}^2+\norm{W}_{\H}^3\right), 
\end{aligned}
\eeq
At last, taking advantage of the estimates \eqref{e:sg-sta} and \eqref{e:Nc} again, we have
\beq
\label{e:DEc3}
\begin{aligned}
   D_{c,3}=& \sup_{t \geq 0}  (1+t)^{\frac{3}{2}} \int_0^t \norm{(1-\chi_{\frac{r_1}{2}})M_{21}(t-s)N_{c}(W(s))}_1 \rmd s\\
    \leq & C\sup_{t \geq 0}  (1+t)^{\frac{3}{2}}\int_0^t \norm{(1-\chi_{\frac{r_1}{2}}) M_{21}(t-s)}_{L^1 \to L^1}\norm{N_{c}(W(s))}_1 \rmd s \\
   \overset{\eqref{e:sg-sta},\;\eqref{e:Nc}}{\leq} & C\left(\norm{W}_{\H}^2+\norm{W}_{\H}^3\right)\sup_{t \geq 0} (1+t)^{\frac{3}{2}} \left(
          \int_0^{t} \rme^{-\lambda_2 (t-s)}(1+s)^{-2} \rmd s \right) \\
    \leq  & C\left(\norm{W}_{\H}^2+\norm{W}_{\H}^3\right)\sup_{t \geq 0} (1+t)^{\frac{3}{2}} \left(
          \rme^{-\frac{\lambda_2t}{2}}\int_0^{t/2} (1+s)^{-2} \rmd s+(1+t/2)^{-2}\int_{t/2}^{t} \rme^{-\lambda_2 (t-s)} \rmd s \right) \\
          \leq &   C \left(\norm{W}_{\H}^2+\norm{W}_{\H}^3\right);\\
         E_{c,3}=& \sup_{t \geq 0} \int_0^t \norm{(1-\chi_{\frac{r_1}{2}})M_{21}(t-s)N_{c}(W(s))}_\infty \rmd s\\
    \leq & C\sup_{t \geq 0}  \int_0^t \norm{(1-\chi_{\frac{r_1}{2}}) M_{21}(t-s)}_{L^\infty \to L^\infty}\norm{N_{c}(W(s))}_\infty \rmd s \\
   \overset{\eqref{e:sg-sta},\;\eqref{e:Nc}}{\leq} & C\left(\norm{W}_{\H}^2+\norm{W}_{\H}^3\right)\sup_{t \geq 0}  \left(
          \int_0^{t} \rme^{-\lambda_2 (t-s)}(1+s)^{-\frac{5}{4}} \rmd s \right) \\
    \leq  & C\left(\norm{W}_{\H}^2+\norm{W}_{\H}^3\right)\sup_{t \geq 0} \left(
          \rme^{-\frac{\lambda_2t}{2}}\int_0^{t/2} (1+s)^{-\frac{5}{4}} \rmd s+(1+t/2)^{-\frac{5}{4}}\int_{t/2}^{t} \rme^{-\lambda_2 (t-s)} \rmd s \right) \\      
   \leq &   C \left(\norm{W}_{\H}^2+\norm{W}_{\H}^3\right).    
\end{aligned}
\eeq
Combining \eqref{e:Ic-int}, \eqref{e:ABCc1}, \eqref{e:ABCc2} and \eqref{e:DEc3}, we conclude that 
\beq\label{e:Ic-tot}
\Rmnum{1}_{c}\leq C\left(\norm{W}_{\H}^2+\norm{W}_{\H}^3\right).
\eeq

\underline{\textit{Estimate of $\Rmnum{1}_s$}} By employing similar arguments as in the proof of \eqref{e:Ic-tot}, we readily exploit \eqref{e:sg-ss},\eqref{e:sg-sta} and \eqref{e:Ns} to conclude that
\beq\label{e:Is-tot}
\Rmnum{1}_{s}\leq C\left(\norm{W}_{\H}^2+\norm{W}_{\H}^3\right).
\eeq
We refer interested readers to Appendix \ref{a:5} for a detailed proof. Combining \eqref{e:T2-gen},\eqref{e:Ic-tot} and \eqref{e:Is-tot}, we conclude the proof of the estimate \eqref{e:T2-est} of $\caT_2(W)$.

To prove \eqref{e:T2-lip}, we note that, similar to \eqref{e:T2-gen} for $\caT_2(W)$, we have
{\footnotesize{\allowdisplaybreaks
\begin{align}\label{e:T2-lip-1}
         &\norm{\caT_2(W_1)-\caT_2(W_2)}_{\H} \nonumber \\
    \leq&\overbrace{\left(\norm{\int_0^tM_{11}(t-s)\Big(N_{c}(W_1)- N_{c}(W_2)\Big)(s)\rmd s}_{\H_c}+\norm{\int_0^tM_{21}(t-s)\Big(N_c(W_1)- N_c(W_2)\Big)(s)\rmd s}_{\H_s} \right) }^{:=\Rmnum{2}_{c}}+\\
    &\overbrace{\left(\norm{\int_0^tM_{12}(t-s)\Big(N_s(W_1)- N_s(W_2)\Big)(s)\rmd s}_{\H_c}+\norm{\int_0^tM_{22}(t-s)\Big(N_{s}(W_1)- N_{s}(W_2)\Big)(s)\rmd s}_{\H_s} \right) }^{:=\Rmnum{2}_{s}} \nonumber
\end{align}}
}
The proof of estimates of $\Rmnum{2}_c$ and $\Rmnum{2}_s$ are exactly the same as the ones for $\Rmnum{1}_c$ and $\Rmnum{1}_s$, except for that we need to replace the estimates \eqref{e:Ncs} of $N_c(W)$ and $N_s(W)$ with the ones of $N_c(W_1)-N_c(W_2)$ and $N_s(W_1)-N_s(W_2)$; that is,
  \begin{equation}\label{e:Ncs-diff}
      \begin{aligned}
          \norm{N_{c}(W_1)-N_c(W_2)}_1  \leq &C (1+t)^{-2}\norm{W_1-W_2}_{\H}\left(\norm{W_1}_{\H}+\norm{W_2}_{\H}+\norm{W_1}_{\H}^2+\norm{W_2}_{\H}^2\right), \\
  \norm{N_{c}(W_1)-N_c(W_2)}_\infty  \leq &C (1+t)^{-\frac{5}{4}}\norm{W_1-W_2}_{\H}\left(\norm{W_1}_{\H}+\norm{W_2}_{\H}+\norm{W_1}_{\H}^2+\norm{W_2}_{\H}^2\right), \\
  \norm{N_{s}(W_1)-N_s(W_2)}_1 \leq &C (1+t)^{-\frac{3}{2}}\norm{W_1-W_2}_{\H}\left(\norm{W_1}_{\H}+\norm{W_2}_{\H}+\norm{W_1}_{\H}^2+\norm{W_2}_{\H}^2\right), \\
  \norm{N_s(W_1)-N_s(W_2)}_\infty  \leq &C (1+t)^{-\frac{3}{4}}\norm{W_1-W_2}_{\H}\left(\norm{W_1}_{\H}+\norm{W_2}_{\H}+\norm{W_1}_{\H}^2+\norm{W_2}_{\H}^2\right).
      \end{aligned}
  \end{equation}
  We are left to show that \eqref{e:Ncs-diff} is true. whose proof is again similar to the one of \eqref{e:Ncs}. 
Therefore, we omit the details except for a brief discussion on the estimate of $N_{c,1}(W_1)-N_{c,1}(W_2)$ for clarity. Noting that 
\[
    \begin{aligned}
        &|N_{c,1}(W_1(t,\bnu)-N_{c,1}(W_2(t,\bnu))|\\=&\left|\int_{\T_1\times\R}k_1(\bnu,\bnut,\bnu-\bnut)\Big(a_1(t,\bnu-\bnut)a_1(t,\bnut)- a_2(t,\bnu-\bnut)a_2(t,\bnut)\Big)\rmd \bnut\right|\\
        \overset{\eqref{e:k1-at0}}{\leq}&C\Big(|a_1|*|\nu_1 (a_1-a_2)| +|a_2|*|\nu_1 (a_1-a_2)|+|\nu_1 a_1|*|(a_1-a_2)|+|\nu_1 a_2|*|(a_1-a_2)|\Big).
    \end{aligned}
\]
we readily conclude
\[
\begin{aligned}
  \norm{N_{c,1}(W_1(t))-N_{c,1}(W_2(t))}_1  \leq &C (1+t)^{-2}(\norm{a_1}_{\H_c}+\norm{a_2}_{\H_c})\norm{a_1-a_2}_{\H_c}; \\
  \norm{N_{c,1}(W_1(t))-N_{c,1}(W_2(t))}_{\infty}\leq &C (1+t)^{-\frac{5}{4}}(\norm{a_1}_{\H_c}+\norm{a_2}_{\H_c})\norm{a_1-a_2}_{\H_c},
\end{aligned}
\]
Lastly, we point out that $N_c(W)$ and $N_s(W)$ consist of quadratic and cubic terms in $W$, and thus are smooth with respect to $W$, yielding
\[
N_{c\backslash s}(W_1)-N_{c\backslash s}(W_2)=\int_0^1 N^\prime_{c\backslash s}(\tau W_1+(1-\tau) W_2)(W_1-W_2)\rmd \tau,
\]
which in turn explains naturally the occurrence of the $\|W_1-W_2\|_{\H}$ term in the estimate \eqref{e:T2-lip}.
\end{proof}

\paragraph{The variation of constant formulation and the contraction mapping $\mathcal{T}$} 

  We aim to show that the map $\mathcal{T}$ is a well-defined contraction in some neighborhood of $\caT_1(W_0)$ in the Banach space $\mathcal{H}$, and thus has a fixed point, which corresponds to a solution to the perturbed Swift-Hohenberg equation \eqref{e:SHE-ivp-v}. Moreover, the $\H$-norm implies the nonlinear stability of the roll solution at the zigzag boundary. More specifically, introducing the notation 
  \[
  B(W,R):=\{V\in\H \mid \|V-W\|_\H\leq R\}, \text{ for any }W\in \H, R>0,
  \]
  we have the following theorem.
\begin{theorem} \label{thm:contraction}
  There exists $\delta > 0$ such that, given $\,\norm{\caT_1(W_0)}_\H < \delta$, the mapping 
  \[
  \caT: B(\caT_1(W_0), \norm{\caT_1(W_0)}_\H)\to B(\caT_1(W_0), \norm{\caT_1(W_0)}_\H)
  \]
  is a well-defined continuous contraction, in the sense that
  \begin{enumerate}
    \item $\mathcal{T}(W) \in B(\caT_1(W_0), \norm{\caT_1(W_0)}_\H)$ for any $W\in B(\caT_1(W_0), \norm{\caT_1(W_0)}_\H)$;
    \item $\norm{\caT(W_1)-\caT(W_2)}_{\H} < \frac{1}{2}\norm{W_1-W_2}_{\H}$ for any $W_1, W_2\in B(\caT_1(W_0), \norm{\caT_1(W_0)}_\H)$.
  \end{enumerate}
\end{theorem}

\begin{proof}
We firstly show that there exists $\delta_1>0$ such that, if $\norm{\caT_1(W_0)}_\H < \delta_1$, then $(\rmnum{1})$ holds. To prove that, we recall the bound given in \eqref{e:T2-est} and have that, for all $W\in\H$,
\beq
\label{e:T-T1}
\norm{\mathcal{T}(W) - \caT_1(W_0)}_\H = \norm{\caT_2(W)}_\H 
    \leq
    C
    \left(
      \norm{W}_\H^2 + \norm{W}_\H^3
    \right).
\eeq
Restricting $W\in B(\caT_1(W_0),
\norm{\caT_1(W_0)}_\H)$, it follows that $\norm{W}_\H \leq 2\norm{\caT_1(W_0)}_\H$, which, together with the above estimate \eqref{e:T-T1} and $\norm{\caT_1(W_0)}_\H < \delta_1$, yields
\[
\begin{aligned}
    \norm{\mathcal{T}(W) - \caT_1(W_0)}_\H \leq & C\left(4\norm{\caT_1(W_0)}_\H + 8\norm{\caT_1(W_0)}_\H^2\right)\norm{\caT_1(W_0)}_\H\\
    \leq & 4C(\delta_1+2\delta_1^2)\norm{\caT_1(W_0)}_\H\\
    \leq &\norm{\caT_1(W_0)}_\H,
\end{aligned}
\]
where the last inequality is true as long as we take 
\beq\label{e:delta1}
0<\delta_1\leq \min\left\{\frac{1}{2}, \frac{1}{8C}\right\},
\eeq
which completes our search for $\delta_1$.

We now show that there exists  $\delta_2>0$ such that, if $\norm{\caT_1(W_0)}_\H < \delta_2$, then $(\rmnum{2})$ holds. Similarly to the search of $\delta_1$ above, we have, for any $W_1, W_2 \in B(\caT_1(W_0),
\norm{\caT_1(W_0)}_\H)$ with $\norm{\caT_1(W_0)}_\H < \delta_2$,
\[
\begin{aligned}
    \norm{\caT(W_1)-\caT(W_2)}_{\H}=& \norm{\caT_2(W_1)-\caT_2(W_2)}_{\H}\\
    \overset{\eqref{e:T2-lip}}{\leq} & C\left(
      \norm{W_1}_\H + \norm{W_2}_\H + \norm{W_1}_\H^2+ \norm{W_2}_\H^2
    \right)\|W_1-W_2\|_{\H}\\
    \leq & 4C(\delta_2+2\delta_2^2)\|W_1-W_2\|_{\H}\\
    \leq &\frac{1}{2}\|W_1-W_2\|_{\H},
\end{aligned}
\]
where the last inequality is true as long as we take 
\beq\label{e:delta2}
0<\delta_2\leq \min\left\{\frac{1}{2}, \frac{1}{16C}\right\},
\eeq
which completes our search for $\delta_2$. We note that the positive constants $C$ in \eqref{e:delta1} and \eqref{e:delta2} are the distinct constants from \eqref{e:T2-est} and \eqref{e:T2-lip} respectively but bear the same notation for conveniences.

Finally, we find that if we choose
$\delta$ such that
\[
  \delta = \min
  \left(\delta_1, \delta_2
  \right),
\]
then both $(\rmnum{1})$ and $(\rmnum{2})$ are true and this concludes the proof of  Theorem \ref{thm:contraction} is proven. 
\end{proof}

We now give the proof of the main theorem.
\begin{proof}[Proof of Theorem \ref{thm:main}]
  From Theorem \ref{thm:contraction} and the estimate \eqref{e:T1-est} of $\caT_1$, for any 
  \[
  \norm{W_0}_1+\norm{W_0}_\infty\leq \delta_0:=\delta/C,
  \]
  where the positive constant $C$ is the one in \eqref{e:T1-est}, we have that $\mathcal{T}$ is a contraction map, which, by Banach's fixed point theorem, gives rise to a unique fixed point 
  \[
  W_*(t,\bnu)=\begin{pmatrix}
      a_*(t,\bnu)\\
      \vsb_*(t,\bnu)
  \end{pmatrix}\in\H,
  \]
  which in turn solves the initial value problem of the perturbed Swift-Hohenberg equation \eqref{e:sys} with the initial condition $W(0,\bnu)=W_0(\bnu)$. Moreover, recalling the mode filter decomposition $T_{mf}$ in \eqref{e:Tmf} and the discrete Bloch-Fourier transform $\caB_d$ in \eqref{e:dbft}, we have that
  \[
  v(t,\x):=\left(\caB_d^{-1}\circ T_{mf}^{-1}W_*\right)(t,\x)=\caB_d^{-1}\left(a_*(t,\bnu)\underline{\widehat{e_c}}(\bnu)+\vsb_*(t,\bnu)\right)
  \]
  solves the perturbed Swift-Hohenberg equation \eqref{e:SHE-ivp-v}
\[
\begin{cases}
   v_t =\caL_p v + \caN_p(v), \\
   v(0,\x)=v_0(\x):=\left(\caB_d^{-1}\circ T_{mf}^{-1}W_0\right)(\x)=\caB_d^{-1}\left(a_0(\bnu)\underline{\widehat{e_c}}(\bnu)+\vsb_0(\bnu)\right).
\end{cases}
\]
As a result, we also have 
{\allowdisplaybreaks
\begin{align*}
    \norm{v(t,\cdot)}_{L^\infty(\R^2)}\leq & \|a_*(t,\cdot)\underline{\widehat{e_c}}(\cdot)+\vsb_*(t,\cdot)\|_{L^1(\T_1\times\R,\ell^1)}\\
    \leq & C\left( \|a_*(t,\cdot)\|_1+\|\vsb_*(t,\cdot)\|_1\right)\\
    \leq & C(1+t)^{-3/4}\norm{W}_\H\\
    \leq & C(1+t)^{-3/4} \norm{\caT W_0}_\H\\
    \leq & C(1+t)^{-3/4} \left(\|W_0\|_1+\|W_0\|_\infty\right)\\
    \leq & C(1+t)^{-3/4} \left(\|\widehat{v_0}\|_{L^1(\R^2)}+\|\widehat{v_0}\|_{L^\infty(\R^2)}\right),
\end{align*}
}
which concludes the proof.
\end{proof}


\appendix

\section{ Proof of Proposition \ref{p:sectorial}: Sectoriality of the linearized operator in discrete Fourier spaces }
\label{a:4}
We recall $\mu_j(\bnu;\kappa)=-(1-(j+\nu_1)^2-\nu_2^2)^2+\kappa^2$ as in \eqref{e:d-lin}, denote $h:=-3u_p^2$, and rewrite the operator in the form
\[
\widehat{\caL_d}=L_0+H,
\]
where 
\[
\begin{matrix}
L_0: & w^{4,p} & \longrightarrow & \ell^p\\
& \underline{u} & \longmapsto & \{\left(\mu_j(\bnu;\kappa)+\veps^2-\kappa^2\right)u_j\}_{j\in\Z},
\end{matrix}
\]
and 
\[
\begin{matrix}
H: & \ell^p & \longrightarrow & \ell^p\\
& \underline{u} & \longmapsto & \underline{\widehat{h}}*\underline{u}.
\end{matrix}
\]
We note that we fix the value of $\bnu$ and thus the dependence of $\widehat{\caL_d}$ and $L_0$ on $\bnu$ is omitted.
We only need to show that the proposition holds for $L_0$ and adding $H$ does not alter these properties. The fact that $L_0$ is closed follows from the fact that the $w^{4,p}$ norm and the graph norm of $L_0$ are equivalent. Noting that $L_0$ is a multiplication operator, we have the spectrum of $L_0$ independent of $p$; that is,
\[
\sigma(L_0)=\{\mu_j(\bnu;\kappa)+\veps^2-\kappa^2\}_{j\in\Z},
\]
and, for any $\lambda\in\rho(L_0)=\C\backslash \sigma(L_0)$, 
\[
\opnorm{(L_0-\lambda)^{-1}}_{\ell^p}\leq \frac{1}{\mathrm{dist}(\lambda, \sigma(L_0))},\quad \text{for any } p\in[1,+\infty].
\]
In addition, $(L_0 - \lambda)^{-1}: \ell^p \to w^{4,p}$ is bounded and the inclusion $w^{4,p} \hookrightarrow \ell^p$ is compact, so
the resolvent of $L_0$ is always compact and thus the spectrum of $L_0$ only consists of eigenvalues.
Denoting $\displaystyle\mu_{max} := \sup_{\bnu\in\T_1\times\R}\max_{j\in\Z}\{\mu_j(\bnu;\kappa)
+ \veps^2-\kappa^2\}$ and fixing $\omega\in(\pi/2,\pi)$, we introduce the sector
\[
S(\mu_{max},\omega):=\left\{\lambda\in\C\;\middle|\; |\arg(\lambda-\mu_{max})|<\omega, \lambda\neq\mu_{max}\right\},
\]
and readily derive that, for any $\lambda\in S(\mu_{max},\omega)$,
\[
\opnorm{(L_0-\lambda)^{-1}}_{\ell^p}\leq \frac{1}{(\sin\omega)|\lambda-\mu_{max}|}, \quad \text{for any } p\in[1,+\infty].
\]
As a result, we conclude that all properties in the proposition hold for $L_0$. 

On the other hand, we note that $H:\ell^p\to\ell^p$ is bounded, uniformly with respect to $p\in[1,\infty]$; that is,
\[
\norm[1]{H\underline{u}}_{\ell^p} \leq \norm[1]{\underline{\widehat{h}}}_{\ell^1}\norm[1]{\underline{u}}_{\ell^p},\quad \text{for any }\underline{u}\in\ell^p, p\in[1,+\infty].
\]
Choosing 
\[
\lambda_0:=\mu_{max}+\frac{2\|\widehat{\underline{h}}\|_{\ell^1}}{\sin\omega},
\]
we have that, for any $\lambda\in S(\lambda_0, \omega):=\left\{\lambda\in\C \;\middle|\; |\arg(\lambda-\lambda_0)|<\omega, \lambda\neq \lambda_0\right\}\subset S(\mu_{max},\omega)$,  
\[
\opnorm{H(L_0-\lambda)^{-1}}_{\ell^p}\leq \opnorm{H}_{\ell^p}\opnorm{(L_0-\lambda)^{-1}}_{\ell^p} \leq \|\widehat{\underline{h}}\|_{\ell^1}\frac{1}{2\|\widehat{\underline{h}}\|_{\ell^1}}=\frac{1}{2}, \quad \text{for any } p\in[1,+\infty],
\]
and thus $\widehat{\caL_d}-\lambda$ is invertible with compact resolvent whose operator norm admits the following estimate.
\[
  \opnorm{(\widehat{\caL_d}-\lambda)^{-1}}_{\ell^p}
  =\opnorm{(L_0 - \lambda)^{-1}\left(I + H(L_0 - \lambda)^{-1}\right)^{-1}}_{\ell^p}\leq \frac{2}{(\sin\omega)|\lambda-\lambda_0|},\quad \text{for any } p\in[1,+\infty].
\]
We are left to show that the spectrum of $\widehat{\caL_d} : \ell^p \to \ell^p$, denoted for now as $\sigma(\widehat{\caL_d},p)$,  is independent of the choice of $p$. For any $p, q\in[1, \infty]$, if $\lambda_*\in \sigma(\widehat{\caL_d},p)$, then $\lambda$ is an eigenvalue for $\widehat{\caL_d} : \ell^p \to \ell^p$ and admits an eigenvector $\underline{u}_* \in \mathcal{D}(\widehat{\caL_d}) = w^{4,p}\subset \ell^q$. Moreover, given any $\lambda\in S(\lambda_0,\omega)$, we have
\[
\underline{u}_*=(\lambda_*-\lambda)(\widehat{\caL_d}-\lambda)^{-1}\underline{u}_*\in w^{4, q},
\]
and thus $\lambda_*\in\sigma(\widehat{\caL_d},q)$ with $\underline{u}_*$ as its eigenfunction. As a result, we have $\sigma(\widehat{\caL_d},p)\subseteq \sigma(\widehat{\caL_d},q)$, for any $p, q\in[1, \infty]$; that is, equivalently, 
$\sigma(\widehat{\caL_d},p)= \sigma(\widehat{\caL_d},q)$, for any $p, q\in[1, \infty]$, which concludes the proof.


\section{ Estimates of $\Rmnum{1}_{s}$ }\label{a:5}
In the proof of the estimate \eqref{e:T2-est} of $\caT_2(W)$ in Proposition \ref{prop:nonlinearbound}, we exploited the estimate \eqref{e:T2-gen}; that is,
\[
\begin{aligned}
    \norm{\caT_2(W)}_{\H}\leq&\overbrace{\left(\norm{\int_0^tM_{11}(t-s)N_{c}(W(s))\rmd s}_{\H_c}+\norm{\int_0^tM_{21}(t-s)N_{c}(W(s))\rmd s}_{\H_s} \right) }^{:=\Rmnum{1}_{c}}+\\
    &\overbrace{\left(\norm{\int_0^tM_{12}(t-s)N_{s}(W(s))\rmd s}_{\H_c}+\norm{\int_0^tM_{22}(t-s)N_{s}(W(s))\rmd s}_{\H_s} \right) }^{:=\Rmnum{1}_{s}},
\end{aligned}
\]
where we discuss the derivation of the estimate \eqref{e:Ic-tot} of $\Rmnum{1}_{c}$ in details. We give the estimates of $\Rmnum{1}_{s}$ in this section. 

\underline{\textit{Estimate of $\Rmnum{1}_{s}$ }} We evaluate $\Rmnum{1}_{s}$ for small and large $\bnu$ respectively; that is,
\beq
\label{e:Is}
\begin{aligned}
  \Rmnum{1}_{s}=&\norm{\int_0^tM_{12}(t-s)N_{s}(W(s))\rmd s}_{\H_c}+\norm{\int_0^tM_{22}(t-s)N_{s}(W(s))\rmd s}_{\H_s} \\
  \leq&\overbrace{\norm{\int_0^t(1-\chi_{\frac{r_1}{2}})M_{12}(t-s)N_{s}(W(s))\rmd s}_{\H_c}}^{:=\Rmnum{1}_{s,1}}+\overbrace{\norm{\int_0^t\chi_{\frac{r_1}{2}}M_{22}(t-s)N_{s}(W(s))\rmd s}_{\H_s}}^{:=\Rmnum{1}_{s,2}}+\\
  &\overbrace{\norm{\int_0^t (1-\chi_{\frac{r_1}{2}})M_{22}(t-s)N_{s}(W(s))\rmd s}_{\H_s}}^{:=\Rmnum{1}_{s,3}}.
\end{aligned}
\eeq
where we use the fact that $\chi_{\frac{r_1}{2}}M_{12}=0$. Moreover, recalling the definition of $\norm{\cdot}_{\H_c}$ and $\norm{\cdot}_{\H_s}$, we have
{\allowdisplaybreaks
\begin{align*}\label{e:Isj}
    \Rmnum{1}_{s,1}\leq&\overbrace{\sup_{t \geq 0} (1+t)^{\frac{3}{4}}\int_0^t \norm{(1-\chi_{\frac{r_1}{2}}) M_{12}(t-s)N_{s}(W(s))}_1 \rmd s}^{:=A_{s,1}}+\overbrace{\sup_{t \geq 0} \int_0^t \norm{(1-\chi_{\frac{r_1}{2}}) M_{12}(t-s)N_{s}(W(s))}_\infty \rmd s}^{:=B_{s,1}}+\\
    &\overbrace{\sup_{t \geq 0}  (1+t)^{\frac{5}{4}}\int_0^t \norm{(1-\chi_{\frac{r_1}{2}}) M_{12}(t-s)N_{s}(W(s))}_1 \rmd s}^{:=C_{s,1}},\\
    \Rmnum{1}_{s,2}\leq &\overbrace{\sup_{t \geq 0} (1+t)^{\frac{3}{2}}\int_0^t \norm{\chi_{\frac{r_1}{2}}M_{22}(t-s)N_{s}(W(s))}_1 \rmd s}^{:=D_{s,2}}+\\
    &\overbrace{\sup_{t \geq 0} \int_0^t \norm{\chi_{\frac{r_1}{2}}M_{22}(t-s)N_{s}(W(s))}_\infty \rmd s}^{:=E_{s,2}},\\
    \Rmnum{1}_{s,3}\leq &\overbrace{\sup_{t \geq 0}  (1+t)^{\frac{3}{2}} \int_0^t \norm{(1-\chi_{\frac{r_1}{2}})M_{22}(t-s)N_{c}(W(s))}_1 \rmd s}^{:=D_{s,3}}+\\
&\overbrace{\sup_{t \geq 0} \int_0^t \norm{(1-\chi_{\frac{r_1}{2}})M_{22}(t-s)N_{s}(W(s))}_\infty \rmd s}^{:=E_{s,3}}.
\end{align*}
}
In other words, we have 
\beq\label{e:Is-int}
\begin{aligned}
    \Rmnum{1}_{s}\leq \sum_{j=1}^3 I_{s,j}\leq A_{s,1}+B_{s,1}+C_{s,1} + \sum_{j=2}^3\left( D_{s,j}+E_{s,j} \right).
\end{aligned}
\eeq

We are left to estimate all the terms in the right hand side of \eqref{e:Is-int}. Taking advantage of the neutral mode estimate \eqref{e:sg-sta} and the estimate \eqref{e:Ns} of $N_s$, we have
{\allowdisplaybreaks
\begin{align}\label{e:ABCs1}
    A_{s,1}=&\sup_{t \geq 0} (1+t)^{\frac{3}{4}} \int_0^t \norm{(1-\chi_{\frac{r_1}{2}})M_{12}(t-s)N_{s}(W(s))}_1 \rmd s \nonumber\\
    \leq & C\sup_{t \geq 0} (1+t)^{\frac{3}{4}}  \int_0^t \norm{(1-\chi_{\frac{r_1}{2}}) M_{12}(t-s)}_{L^1 \to L^1}\norm{N_{s}(W(s))}_1 \rmd s \nonumber\\
   \overset{\eqref{e:sg-sta},\;\eqref{e:Ns}}{\leq} & C\left(\norm{W}_{\H}^2+\norm{W}_{\H}^3\right)\sup_{t \geq 0} (1+t)^{\frac{3}{4}}  \left(
          \int_0^{t} \rme^{-\lambda_2 (t-s)}(1+s)^{-\frac{3}{2}} \rmd s \right) \nonumber\\
    \leq  & C\left(\norm{W}_{\H}^2+\norm{W}_{\H}^3\right)\sup_{t \geq 0} (1+t)^{\frac{3}{4}} \left(
          \rme^{-\frac{\lambda_2t}{2}}\int_0^{t/2} (1+s)^{-\frac{3}{2}} \rmd s+(1+t/2)^{-\frac{3}{2}}\int_{t/2}^{t} \rme^{-\lambda_2 (t-s)} \rmd s \right) \nonumber\\    
   \leq &   C \left(\norm{W}_{\H}^2+\norm{W}_{\H}^3\right) ; \nonumber\\
B_{s,1}=&\sup_{t \geq 0} \int_0^t \norm{(1-\chi_{\frac{r_1}{2}}) M_{12}(t-s)N_{s}(W(s))}_\infty \rmd s \nonumber\\
 \leq&\sup_{t \geq 0}\int_0^{t} \norm{(1-\chi_{\frac{r_1}{2}}) M_{12}(t-s)}_{L^\infty \to L^\infty}
          \norm{N_{s}(W(s))}_\infty
          \rmd s \nonumber\\
         \overset{\eqref{e:sg-sta},\;\eqref{e:Ns}}{\leq} & C\left(\norm{W}_{\H}^2+\norm{W}_{\H}^3\right)\sup_{t \geq 0}  \left(
          \int_0^{t} \rme^{-\lambda_2 (t-s)}(1+s)^{-\frac{3}{4}} \rmd s \right) \\
    \leq  & C\left(\norm{W}_{\H}^2+\norm{W}_{\H}^3\right)\sup_{t \geq 0}  \left(
          \rme^{-\frac{\lambda_2t}{2}}\int_0^{t/2} (1+s)^{-\frac{3}{4}} \rmd s+(1+t/2)^{-\frac{3}{4}}\int_{t/2}^{t} \rme^{-\lambda_2 (t-s)} \rmd s \right) \nonumber\\     
   \leq &   C \left(\norm{W}_{\H}^2+\norm{W}_{\H}^3\right) ; \nonumber\\
 C_{s,1}=&\sup_{t \geq 0}  (1+t)^{\frac{5}{4}}\int_0^t \norm{\nu_1(1-\chi_{\frac{r_1}{2}}) M_{12}(t-s)N_{c}(W(s))}_1 \rmd s \nonumber\\
 \leq& C\sup_{t \geq 0} (1+t)^{\frac{5}{4}}  \int_0^t \norm{\nu_1(1-\chi_{\frac{r_1}{2}}) M_{12}(t-s)}_{L^1 \to L^1}\norm{N_{s}(W(s))}_1 \rmd s \nonumber\\
   \overset{\eqref{e:sg-sta},\;\eqref{e:Ns}}{\leq} & C\left(\norm{W}_{\H}^2+\norm{W}_{\H}^3\right)\sup_{t \geq 0} (1+t)^{\frac{5}{4}}  \left(
          \int_0^{t} \rme^{-\lambda_2 (t-s)}(1+s)^{-\frac{3}{2}} \rmd s \right) \nonumber\\
    \leq  & C\left(\norm{W}_{\H}^2+\norm{W}_{\H}^3\right)\sup_{t \geq 0} (1+t)^{\frac{5}{4}} \left(
          \rme^{-\frac{\lambda_2t}{2}}\int_0^{t/2} (1+s)^{-\frac{3}{2}} \rmd s+(1+t/2)^{-\frac{3}{2}}\int_{t/2}^{t} \rme^{-\lambda_2 (t-s)} \rmd s \right) \nonumber\\     
   \leq &   C \left(\norm{W}_{\H}^2+\norm{W}_{\H}^3\right) , \nonumber\\
\end{align}
}
Similarly, taking advantage of the estimates \eqref{e:sg-ss} and \eqref{e:Ns}, we have
\beq
\label{e:ABCs2}
\begin{aligned}
    D_{s,2}=&\sup_{t \geq 0} (1+t)^{\frac{3}{2}} \int_0^t \norm{\chi_{\frac{r_1}{2}}M_{22}(t-s)N_{s}(W(s))}_1 \rmd s\\
    \leq & C\sup_{t \geq 0} (1+t)^{\frac{3}{2}}  \int_0^t \norm{\chi_{\frac{r_1}{2}} M_{22}(t-s)}_{L^1 \to L^1}\norm{N_{s}(W(s))}_1 \rmd s \\
   \overset{\eqref{e:sg-ss},\;\eqref{e:Ns}}{\leq} & C\left(\norm{W}_{\H}^2+\norm{W}_{\H}^3\right)\sup_{t \geq 0} (1+t)^{\frac{3}{2}}  \left(
          \int_0^{t} \rme^{-\lambda_1 (t-s)}(1+s)^{-\frac{3}{2}} \rmd s \right) \\
    \leq  & C\left(\norm{W}_{\H}^2+\norm{W}_{\H}^3\right)\sup_{t \geq 0} (1+t)^{\frac{3}{2}} \left(
          \rme^{-\frac{\lambda_1 t}{2}}\int_0^{t/2} (1+s)^{-\frac{3}{2}} \rmd s+(1+t/2)^{-\frac{3}{2}}\int_{t/2}^{t} \rme^{-\lambda_1 (t-s)} \rmd s \right) \\      
   \leq &   C \left(\norm{W}_{\H}^2+\norm{W}_{\H}^3\right) ;\\
   E_{s,2}=&\sup_{t \geq 0} \int_0^t \norm{\chi_{\frac{r_1}{2}} M_{22}(t-s)N_{s}(W(s))}_\infty \rmd s \\
 \leq&\sup_{t \geq 0}\int_0^{t} \norm{\chi_{\frac{r_1}{2}}M_{22}(t-s)}_{L^\infty \to L^\infty}
          \norm{N_{s}(W(s))}_\infty
          \rmd s \\
         \overset{\eqref{e:sg-ss},\;\eqref{e:Ns}}{\leq} & C\left(\norm{W}_{\H}^2+\norm{W}_{\H}^3\right)\sup_{t \geq 0}  \left(
          \int_0^{t} \rme^{-\lambda_1(t-s)}(1+s)^{-\frac{3}{4}} \rmd s \right) \\
    \leq  & C\left(\norm{W}_{\H}^2+\norm{W}_{\H}^3\right)\sup_{t \geq 0}  \left(
          \rme^{-\frac{\lambda_1 t}{2}}\int_0^{t/2} (1+s)^{-\frac{3}{4}} \rmd s+(1+t/2)^{-\frac{3}{4}}\int_{t/2}^{t} \rme^{-\lambda_1 (t-s)} \rmd s \right) \\      
   \leq &   C \left(\norm{W}_{\H}^2+\norm{W}_{\H}^3\right) .
\end{aligned}
\eeq
At last, taking advantage of the estimates \eqref{e:sg-sta} and \eqref{e:Ns} again, we have
\beq
\label{e:DEs3}
\begin{aligned}
   D_{s,3}=& \sup_{t \geq 0}  (1+t)^{\frac{3}{2}} \int_0^t \norm{(1-\chi_{\frac{r_1}{2}})M_{22}(t-s)N_{s}(W(s))}_1 \rmd s\\
    \leq & C\sup_{t \geq 0}  (1+t)^{\frac{3}{2}}\int_0^t \norm{(1-\chi_{\frac{r_1}{2}}) M_{22}(t-s)}_{L^1 \to L^1}\norm{N_{s}(W(s))}_1 \rmd s \\
   \overset{\eqref{e:sg-sta},\;\eqref{e:Ns}}{\leq} & C\left(\norm{W}_{\H}^2+\norm{W}_{\H}^3\right)\sup_{t \geq 0} (1+t)^{\frac{3}{2}} \left(
          \int_0^{t} \rme^{-\lambda_2 (t-s)}(1+s)^{-\frac{3}{2}} \rmd s \right) \\
    \leq  & C\left(\norm{W}_{\H}^2+\norm{W}_{\H}^3\right)\sup_{t \geq 0} (1+t)^{\frac{3}{2}} \left(
          \rme^{-\frac{\lambda_2t}{2}}\int_0^{t/2} (1+s)^{-\frac{3}{2}} \rmd s+(1+t/2)^{-\frac{3}{2}}\int_{t/2}^{t} \rme^{-\lambda_2 (t-s)} \rmd s \right) \\
          \leq &   C \left(\norm{W}_{\H}^2+\norm{W}_{\H}^3\right);\\
E_{s,3}=& \sup_{t \geq 0} \int_0^t \norm{(1-\chi_{\frac{r_1}{2}})M_{22}(t-s)N_{s}(W(s))}_\infty \rmd s\\
    \leq & C\sup_{t \geq 0}  \int_0^t \norm{(1-\chi_{\frac{r_1}{2}}) M_{22}(t-s)}_{L^\infty \to L^\infty}\norm{N_{s}(W(s))}_\infty \rmd s \\
   \overset{\eqref{e:sg-ss},\;\eqref{e:Ns}}{\leq} & C\left(\norm{W}_{\H}^2+\norm{W}_{\H}^3\right)\sup_{t \geq 0}  \left(
          \int_0^{t} \rme^{-\lambda_2 (t-s)}(1+s)^{-\frac{3}{4}} \rmd s \right) \\
    \leq  & C\left(\norm{W}_{\H}^2+\norm{W}_{\H}^3\right)\sup_{t \geq 0} \left(
          \rme^{-\frac{\lambda_2t}{2}}\int_0^{t/2} (1+s)^{-\frac{3}{4}} \rmd s+(1+t/2)^{-\frac{3}{4}}\int_{t/2}^{t} \rme^{-\lambda_2 (t-s)} \rmd s \right) \\      
   \leq &   C \left(\norm{W}_{\H}^2+\norm{W}_{\H}^3\right).    
\end{aligned}
\eeq
Combining \eqref{e:Is-int}, \eqref{e:ABCs1}, \eqref{e:ABCs2} and \eqref{e:DEs3}, we conclude that 
\beq\label{e:Is-tot}
\Rmnum{1}_{s}\leq C\left(\norm{W}_{\H}^2+\norm{W}_{\H}^3\right).
\eeq

\section{Declarations of Funding and Conflicts of interests}

This work was supported by the National Science Foundation through grant DMS-1815079. The authors have no other relevant financial or non-financial interests to disclose.

\printbibliography

\end{document}